\documentclass[aps,prd,eqsecnum,preprint,groupedaddress,longbibliography,nofootinbib]{revtex4}

\usepackage{graphicx}
\usepackage{amsmath}
\usepackage{amssymb}
\usepackage[mathscr]{euscript}

\newcommand{\dS}[0]{\textrm{dS}_{k=0}}
\newcommand{\CVPS}[0]{\Gamma(\dS)}
\newcommand{\dSh}[0]{\textrm{dS}_{k=-1}}
\newcommand{\CVPSh}[0]{\Gamma(\dSh)}

\newcommand{\order}[2]{\mathcal{O}(#1^{#2})}

\newcommand{\csch}{\mathrm{csch} \,}

\begin{document}

\title{Phase Spaces for asymptotically de Sitter Cosmologies}
\author{William R. Kelly}
\email{wkelly@physics.ucsb.edu}
\affiliation{University of California at Santa Barbara, Santa Barbara, CA 93106, USA}
\author{Donald Marolf}
\email{marolf@physics.ucsb.edu}
\affiliation{University of California at Santa Barbara, Santa Barbara, CA 93106, USA}


\begin{abstract}
We construct two types of phase spaces for asymptotically de Sitter Einstein-Hilbert gravity in each spacetime dimension $d \ge 3$.  One type contains solutions asymptotic to the expanding spatially-flat ($k=0$) cosmological patch of de Sitter space while the other is asymptotic to the expanding hyperbolic $(k=-1)$ patch.  Each phase space has a non-trivial asymptotic symmetry group (ASG) which includes the isometry group of the corresponding de Sitter patch.  For $d=3$ and $k=-1$ our ASG also contains additional generators and leads to a Virasoro algebra with vanishing central charge.  Furthermore, we identify an interesting algebra (even larger than the ASG) containing two Virasoro algebras related by a reality condition and having imaginary central charges $\pm i \frac{3\ell}{2G}$.  Our charges agree with those obtained previously using dS/CFT methods for the same asymptotic Killing fields showing that (at least some of) the dS/CFT charges act on a well-defined phase space.   Along the way we show that, despite the lack of local degrees of freedom, the $d=3, k=-1$ phase space is non-trivial even in pure $\Lambda > 0$ Einstein-Hilbert gravity due to the existence of a family of `wormhole' solutions labeled by their angular momentum, a mass-like parameter $\theta_0$, the topology of future infinity ($I^+$), and perhaps additional internal moduli.  These solutions are $\Lambda > 0$ analogues of BTZ black holes and exhibit a corresponding mass gap relative to empty de Sitter.
\end{abstract}

\maketitle

\section{Introduction}
Spacetimes that approximate de Sitter space (dS) form the basis of inflationary early universe cosmology and also give a rough description of our current universe.  One expects this description to further improve in the future as the cosmological expansion dilutes the various forms of matter, and that in tens of Gyrs it will become quite good indeed.   Yet certain classic issues in gravitational physics, such as the construction of phase spaces and conserved charges, are less well developed in the de Sitter context than  for asymptotically flat or asymptotically AdS spacetimes; see e.g.
\cite{ADM,ReggeTeitelboim,HenneauxTeitelboimAsympAdS,BrownHenneauxCentralCharge,AshtekarCVPS,CovarPhaseSpace}.
While there have been many discussions of de Sitter charges (see
\cite{CCStability,Banados:1998tb,Park:1998qk,Strominger:2001pn,balasubramaniandS,Shiromizu:2001bg,KastorCKV,JagerPhD,LuoPositiveMass,StromingerASG}) over a broad span of time,
most of these \cite{CCStability,Banados:1998tb,Park:1998qk,Strominger:2001pn,balasubramaniandS,JagerPhD,StromingerASG}  do not construct a phase space in which the charges generate the associated diffeomorphisms while the remainder \cite{Shiromizu:2001bg,KastorCKV,LuoPositiveMass}  define phase spaces in which many of the expected charges diverge.  In particular, since~\cite{Strominger:2001pn,balasubramaniandS,JagerPhD,StromingerASG} fix the induced metric on future infinity the symplectic structure necessarily vanishes.  We shall impose no such condition here.

This hole in the literature is presumably due, at least in part, to the fact that global de Sitter space admits a compact Cauchy surface of topology $S^{d-1}$. It is thus natural to define a phase space (which we call the $k=+1$ phase space, $\Gamma({\textrm{dS}_{k=+1}})$) which contains {\it all} solutions  with an $S^{d-1}$ Cauchy surface.   Since $S^{d-1}$ is compact, there is no need to impose further boundary conditions.  The constraints then imply that  all gravitational charges vanish identically.  All diffeomorphisms are gauge symmetries and the asymptotic symmetry group is trivial.

\begin{figure}
\includegraphics[width=4.5in]{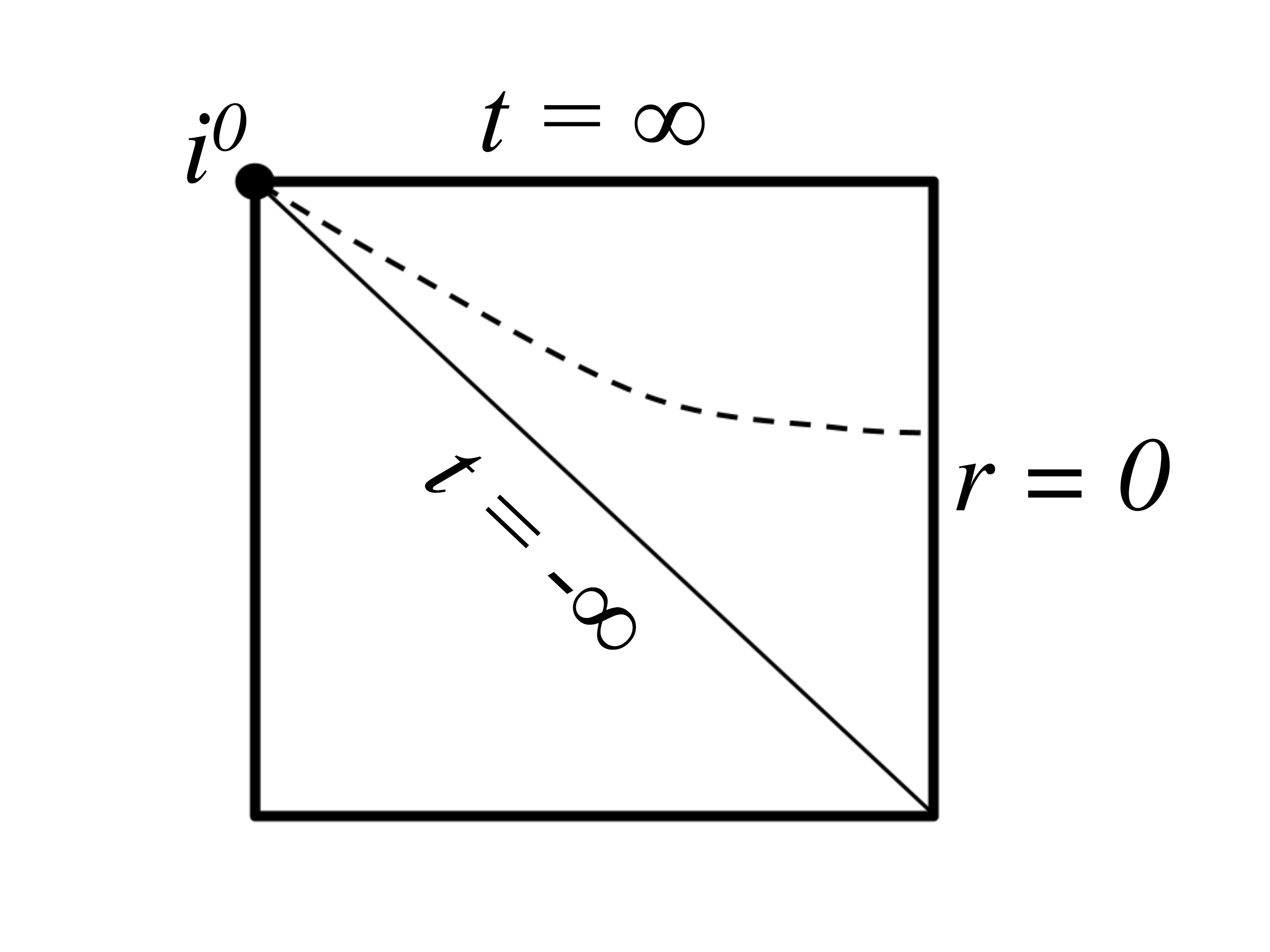}
\caption{A conformal diagram of dS${}_d$.  The region above the diagonal line is the $k=0$ cosmological patch.  Each point represents an $S^{d-2}$.  Our boundary conditions are applied at the $S^{d-2}$ labeled $i^0$, which we call spatial infinity.  The dashed line shows a representative $(t= {\rm constant})$ slice.}
\label{conformalDiagram}
\end{figure}

On the other hand, it is natural in cosmological contexts to consider pieces of de Sitter space which may be foliated by either flat ($k=0$) or hyperbolic ($k=-1$) Cauchy surfaces. We call these patches $\dS$ and $\dSh$ respectively, see  Figs.~\ref{conformalDiagram}, \ref{conformalDiagramHyp}.   These Cauchy surfaces are non-compact, and boundary conditions are required in the resulting asymptotic regions.  The purpose of this paper is to construct associated phase spaces (for both $k=0,-1$) of asymptotically de Sitter solutions in $d \ge 3$ spacetime dimensions for which the expected charges are finite and conserved.
As there are claims \cite{Kleppe:1993fz,Miao:2009hb,Miao:2010vs,Miao:2011fc,Miao:2011ng,Kahya:2011sy}  in the literature that the so-called `dilatation symmetry' of the $k=0$ patch is broken at the quantum level (though see \cite{Allen:1986ta,Allen:1986dd,Allen:1986tt,Higuchi:1986py,Higuchi:1991tn,Higuchi:1991tk,Higuchi:2000ye,Higuchi:2001uv,Higuchi:2011vw}), it is particularly important to verify that this is indeed a symmetry of an appropriate classical gravitational phase space for $k=0$.

In most cases below the resulting asymptotic symmetry group (ASG) is the isometry group of the associated (flat- or hyperbolic-sliced) patch of dS, though for $k=-1$ and $d=3$ we find that the obvious rotational symmetry is enlarged to a (single) Virasoro algebra in the ASG.  The structure is somewhat similar to that recently seen in the Kerr/CFT context \cite{Guica:2008mu,Bredberg:2011hp}, though in our present case the central charge vanishes due to a reflection symmetry in the angular direction.  We also identify an interesting algebra somewhat larger than the ASG which contains two Virasoro algebras related by a reality condition and having imaginary central charges (in agreement with \cite{Fjelstad:2002wf,Balasubramanian:2002zh,Maldacena:2002vr,Ouyang:2011fs}).  We note, however, that the extra generators (outside the ASG) have incomplete flows on our classical phase space. Thus real classical charges of this sort will not lead to self-adjoint operators at the quantum level.

We construct the phase spaces $\CVPS$ and $\CVPSh$ associated with $\dS$ and $\dSh$ below in sections \ref{k=0} and \ref{k=-1}. In each case, we find that our final expressions agree with the relevant charges of \cite{Strominger:2001pn,balasubramaniandS,StromingerASG}\footnote{We expect the same to be true of \cite {Banados:1998tb,Park:1998qk}.  However, the fact that \cite{Banados:1998tb,Park:1998qk} used a Chern-Simons formulation makes direct comparison non-trivial; we will not attempt it here. We also make no direct comparison with ref. \cite{CCStability}, which worked perturbatively around dS, though we again expect agreement in the appropriate regime.}.  This agreement provides a sense in which those charges generate canonical transformations on a well-defined phase space.  

The case $d=3, k=-1$ merits special treatment.  Despite the lack of local degrees of freedom, section \ref{wormholes} shows the phase space to be non-trivial due to a class of de Sitter wormholes.  These solutions are $\Lambda > 0$ analogues of BTZ black holes and exhibit a corresponding mass gap.  We close with some final discussion in section \ref{disc} which in particular compares our phase space with those of \cite{Shiromizu:2001bg,KastorCKV,LuoPositiveMass}.

\section{The phase space $\CVPS$}
\label{k=0}

Our first phase space
will consist of spacetimes asymptotic to the expanding spatially-flat ($k=0$) patch of de Sitter space (see figure \ref{conformalDiagram}) in $d\ge 3$ spacetime dimensions for which the metric takes the familiar form
\begin{equation}
\label{k=0dS}
ds^2 = - dt^2 + e^{2t/\ell} \delta_{ij} dx^i dx^j.
\end{equation}
Here $\delta_{ij}$ is a Kroenecker delta and $i,j$ range over the $d-1$ spatial coordinates.
We will refer to this patch as $\dS$ and the corresponding phase space as $\CVPS$.  For future reference we note that the symmetries of $\dS$ are generated by three types of Killing fields (dilations, translations, and rotations) which take the following forms

\begin{eqnarray}
\label{k=0KVFs}
{\rm Dilations:} &  \xi_D^a & =  (\partial_t)^a - x^a/\ell ,\cr
{\rm Translations:} & \xi_{P^i}^a & = (\partial_i)^a, \cr
{\rm Rotations:} & \xi_{L^{ij}}^a & = (2x_{[i}\partial_{j]})^a.
\end{eqnarray}

Elements of our phase space are globally hyperbolic solutions to the Einstein equation in $d\ge3$ spacetime dimensions with positive cosmological constant
\begin{eqnarray}
\Lambda = \frac{(d-1)(d-2)}{2\ell^2}
\end{eqnarray}
and topology ${\mathbb R}^d.$  Introducing a time-function $t$ defines a foliation of ($t={\rm constant}$) spacelike slices $\Sigma$.  Choosing coordinates $x^i \in {\mathbb R}^{d-1}$ on each slice, the metric may then be written in the form
\begin{eqnarray}
ds^2 = -N^2 dt^2 + h_{ij} (dx^i + N^i dt)(dx^j + N^j dt). \label{dSmetric}
\end{eqnarray}
We define $\CVPS$ to contain such spacetimes for which, on any $t = (\textrm{constant})$ slice $\Sigma$, the induced metric $h_{ij}$, the canonical momentum $\tilde{\pi}^{ij} = \sqrt{h} \pi^{ij}$, the lapse $N$, and the shift $N^i$ satisfy the boundary conditions\footnote{A study of the symplectic structure (eq. \eqref{SymplecticStructure} below) indicates that these boundary conditions can be significantly relaxed, presumably allowing radiation that falls off more slowly at large $r$. However, doing so requires non-trivial use of the equations of motion to make explicit the fact that the charges associated with \eqref{k=0KVFs} are finite.  We have not attempted to complete such an analysis as we see no obvious advantage to weakening the boundary conditions \eqref{BoundaryConditions}.}
\begin{align} \label{BoundaryConditions}
\Delta h_{ij} &= h_{ij}^{(d-1)} + \order{r}{-(d-1+\epsilon)}, \cr
\Delta{\pi}^{ij} &= \pi^{ij}_{(d-2)} + \pi^{ij}_{(d-1)} + \order{r}{-(d-1+\epsilon)} \cr
N  &= 1+ N^{(d-2)} + \order{r}{-(d-1)} \cr
N^i &= N^i_{(d-3)} + \order{r}{-(d-2)},
\end{align}
at large $r = \sqrt{\delta_{ij} x^i x^j}$, with
\begin{subequations}
\label{subh}
\begin{align}
h_{ij}^{(d-1)} =  \frac{(\textrm{Any function of $\Omega$})_{ij}}{r^{d-1}}, \ \ & \\
\pi^{ij}_{(d-2)} = \frac{(\textrm{Odd function of $\Omega$})^{ij}}{r^{d-2}}, \ \ & \ \
\pi^{ij}_{(d-1)} = \frac{(\textrm{Any function of $\Omega$})^{ij}}{r^{d-1}} \\
N^{(d-2)} =  \frac{(\textrm{Odd function of $\Omega$})}{r^{d-2}}, \ \ & \ \ N^i_{(d-3)} = \frac{(\textrm{Even function of $\Omega$})^i}{r^{d-3}}
\end{align}
\end{subequations}
where ${\Omega}$ denotes a set of angular coordinates on $r=(\textrm{constant})$ slices and where
\begin{align} \label{Deltahandpi}
\Delta h_{ij} = h_{ij} -  e^{2t/\ell}\delta_{ij} = h_{ij} - \bar{h}_{ij}, \ \ \ \
\Delta \pi^{ij} = \pi^{ij} + \frac{(d-2) }{\ell}e^{-2 t/\ell}\delta^{ij} = \pi^{ij} - \bar{\pi}^{ij},
\end{align}
where $\bar{h}_{ij}$ and $\bar{\pi}^{ij}$ are the induced metric and momentum associated with~\eqref{k=0dS} (in general overbars will denote quantities associated with~\eqref{k=0dS}).  We emphasize that, since the above conditions restrict the behavior only in the limit $r \rightarrow \infty$, they in no way restrict the induced (conformal) metric on future infinity at any finite $r$.

In order for time evolution to preserve~\eqref{BoundaryConditions} the lapse, shift and momentum must satisfy the additional relation
\begin{align} \label{lapseshiftcondition}
\pi_{ij}^{(d-2)} + \partial_{(i} N_{j)}^{(d-3)} + N^{(d-2)} \frac{\bar h_{ij}}{\ell} = 0.
\end{align}
This final condition was obtained by writing the equations of motion to leading order, imposing~\eqref{BoundaryConditions} and requiring that  $\dot h^{(d-2)}_{ij}=0$ (no further condition is required to make $\dot \pi_{(d-2)}^{ij}$ odd).   In all of the explicit examples we consider below this condition is satisfied trivially.  We also assume that the $n$th derivative of the $\order{r}{-(d-1+\epsilon)}$ term in \eqref{BoundaryConditions} is $\order{r}{-(d-1+n+\epsilon)}$.

The definitions~\eqref{Deltahandpi} were chosen so that $\Delta h_{ij} = 0 = \Delta{\pi}^{ij}$ for exact de Sitter (Eq.~\eqref{k=0dS}).  We also note that~\eqref{BoundaryConditions} together with the constraints~\eqref{constraints}, ensures that
\begin{align} \label{pitildeBC}
\Delta{\tilde\pi}^{ij} &= \frac{(\textrm{Odd function of $\Omega$})^{ij}}{r^{d-2}} + \mathcal{O}\left(r^{-(d-1)}\right),
\end{align}
with
\begin{align}
\Delta{\tilde\pi}^{ij} = \sqrt{h}\pi^{ij} - \sqrt{\bar h} \bar\pi^{ij}.
\end{align}

Let us now consider two tangent vectors $(\delta_1 h_{ij},\delta_1\tilde{\pi}^{ij})$ and $(\delta_2 h_{ij},\delta_2\tilde{\pi}^{ij})$ to $\CVPS$.  In order for our phase space to be well defined we must show the symplectic product of these two tangent vectors to be finite and independent of the Cauchy surface on which it is evaluated, i.e. independent of $t$.
Our boundary conditions suffice to guarantee both of these conditions.  Equations~\eqref{BoundaryConditions} and~\eqref{pitildeBC} imply convergence of the standard expression
\begin{eqnarray}
\omega(\delta_1 g,\delta_2 g) = \frac{1}{4\kappa} \int_\Sigma \left(\delta_1 h_{ij}\delta_2\tilde{\pi}^{ij} - \delta_2 h_{ij}\delta_1\tilde{\pi}^{ij}\right) \label{SymplecticStructure}
\end{eqnarray}
for the symplectic product (see e.g. \cite{PalmerSymplecticForm}).  Furthermore, we will show in section \ref{CC} below that the (time-depenedent) Hamiltonian $H(\partial_t)$ (see~\eqref{Hz}) defined by some $N,N^i$ satisfying \eqref{BoundaryConditions} i) has well-defined variations and ii) generates an evolution that preserves the boundary conditions \eqref{BoundaryConditions} on $h_{ij}$ and $\tilde \pi^{ij}$.  This in turn guarantees that $\omega(\delta_1 g,\delta_2 g)$ is time independent.\footnote{A finite well-defined Hamiltonian that preserves the phases space ensures that the Poisson bracket is conserved.  Since the symplectic product is the inverse of the Poisson structure, it too must be conserved.\label{symfoot}}

Thus, we conclude that $\CVPS$ is well-defined.  Below we compute asymptotic symmetries and conserved charges, largely following the approach of~\cite{ReggeTeitelboim,HenneauxTeitelboimAsympAdS,BrownHenneauxCentralCharge}.

\subsection{Asymptotic Symmetries}

We begin by using the fact that linearized diffeomorphisms generated by any element of our ASG must map~\eqref{k=0dS} onto a solution satisfying~\eqref{BoundaryConditions}.  Consider the metric $\bar h_{ij}$ induced on a $t=(\textrm{constant})$ slice of~\eqref{k=0dS} and its pullback into the bulk spacetime which we call $\bar h_{ab}$.  We also introduce $h^a_i$ which is the projector from the spacetime onto $\Sigma$.  If $\xi$ is in our ASG then
\begin{eqnarray}
\delta_\xi \bar h_{ij} &\equiv& h^a_i h^b_j \pounds_\xi \bar h_{ab} \cr
&=&  h^a_i h^b_j \left[\xi^c \bar\nabla_c (\bar h_{ab}) + 2 \bar h_{c(a}\bar\nabla_{b)} \xi^{c}\right] \cr
&=& 2 h^a_i h^b_j \bar\nabla_{(a} \xi_{b)}  , \label{deltazetah}
\end{eqnarray}
must vanish as $r\rightarrow\infty$ fast enough so that $\bar h_{ij} + \delta_\xi \bar h_{ij}$ satisfies~\eqref{BoundaryConditions}.  So,  up to terms which vanish at $r\rightarrow\infty$, $\xi$ must satisfy
\begin{align}
\label{ckvf}
\partial_{(i} \vec\xi_{j)} = \frac{\xi_\perp e^{2t/\ell}}{\ell}\delta_{ij},
\end{align}
where we have defined the tangential and normal parts $\xi_\perp, \vec \xi^a$ to $\Sigma$ via the decomposition
\begin{equation}
\label{parts}
\xi^a = \xi_\perp n^a + \vec\xi^a.
\end{equation}
Note that \eqref{ckvf} is the conformal Killing equation for vectors $\vec\xi^i$ in Euclidean $\mathbb{R}^{d-1}$ with confromal factor $\xi_\perp e^{2t/\ell}/\ell$.

We wish to discard solutions to this equation which are either pure gauge ($\omega(\pounds_\xi g,\delta g)=0$) or do not preserve our boundary conditions.  It is shown at the end of appendix \ref{finiteQ} that if $\xi$ vanishes as $r\rightarrow\infty$ then $\omega(\pounds_\xi g,\delta g)=0$.  For $d>3$ the remaining solutions are the conformal group of $\mathbb{R}^{d-1}$.  We find using~\eqref{LieDhpi} that our boundary conditions \eqref{BoundaryConditions} are not invariant under special conformal transformations\footnote{Specfically, acting on the Schwarzschild de Sitter solution \eqref{SdS2} below twice with a Lie derivative along the generator of special conformal transformations gives a term which violates our boundary conditions.}.  Excluding such transformations leaves a group isomorphic to the isometries of $\dS$ (see \eqref{k=0KVFs}).

Due to the infinite-dimensional conformal group of the plane there are additional solutions to \eqref{ckvf} for $d=3$, each of which can be described by a potential satisfying Laplace's equation.  Expanding this potential in sperical harmonics we obtain a set of symmetries which fall off with various powers of $r$.  Vector fields with terms of order $r^2$ or higher violate our boundary conditions while those of order $r^{-1}$ or lower are pure gauge because they vanish at infinity.  What remains are four vector fields (two of order $r^1$, two of order $r^0$) corresponding to the four isometries of $\dS$ for $d=3$.

We now show that our phase space is closed under the action of the expected symmetry group~\eqref{k=0KVFs}.  To do so, we consider an arbitrary solution $(h_{ij},\tilde \pi^{ij})$ satisfying~\eqref{BoundaryConditions} and show that $(h_{ij} + \delta_\xi h_{ij},\tilde \pi^{ij} + \delta_\xi \tilde \pi^{ij})$ also satisfies~\eqref{BoundaryConditions} where $\xi$ is one of the vector fields~\eqref{k=0KVFs}.

First consider a purely spatial vector $\xi$, i.e. a translation or rotation.  From the expressions
\begin{eqnarray}
\pounds_{\xi}h_{ij} &=& \vec\xi^k \partial_k \Delta h_{ij} + 2 \Delta h_{k(i} \partial_{j)} \vec\xi^k  \cr
\pounds_{\xi}{\pi}^{ij} &=& \vec\xi^k \partial_k \Delta\pi^{ij} -2 \Delta \pi^{k(i} \partial_k \vec\xi^{j)}, \label{LieDhpi}
\end{eqnarray}
we can see that our boundary conditions are preserved by diffeomorphisms along these vector fields.

To see that $\xi_D$ preserves our boundary conditions note that
\begin{align}
\pounds_{\xi_D} = \pounds_{t} - \pounds_{x/\ell}.
\end{align}
Together with the canonical equations of motion, the boundary conditions \eqref{BoundaryConditions} and \eqref{lapseshiftcondition} ensure that $\pounds_{t}$ preserves $\CVPS$, and it is straightforward to verify that $\pounds_{x/\ell}$ does as well using~\eqref{LieDhpi}.  Thus, our boundary conditions are also preserved by $\xi_D$.  This completes our proof that the asymptotic symmetry group of $\CVPS$ is given by the isometries of $\dS$.

\subsection{Conserved Charges}
\label{CC}

Our next task is to construct a corresponding set of conserved charges.
As described by Regge and Teitelboim~\cite{ReggeTeitelboim}, the fact that any such charge $H(\xi)$ must generate diffeomorphisms along $\xi$ implies that $H(\xi)$  is a linear combination of the gravitational constraints determined by the relevant vector field $\xi$, together with certain surface terms chosen to ensure that the charges have well-defined variations with respect to $h_{ij}$ and $\tilde \pi^{ij}$.    So long as the boundary conditions are sufficiently strong, the result takes the standard form
\begin{align}
H(\xi) &= \frac{1}{2\kappa} \int_\Sigma \sqrt{h} \left(\xi_\perp {\cal H} + \vec \xi^i {\cal H}_i   \right) + \frac{1}{\kappa} \int_{\partial\Sigma} (dr)_i \vec \xi^j \left(\Delta\tilde\pi^{ik} h_{kj} + \tilde\pi^{ik} \Delta h_{kj}  - \frac{\tilde\pi^{kl}\Delta h_{kl} }{2}{\delta^i}_j \right) \nonumber \\
&+ \frac{1}{2\kappa}\int_{\partial\Sigma}  \sqrt{\sigma} \hat{r}_l G^{ijkl} \left(\xi_\perp D_k  \Delta h_{ij} - \Delta h_{ij} D_k \xi_\perp\right), \label{Hz}
\end{align}
in terms of the Hamiltonian and momentum constraints
\begin{eqnarray}
\label{constraints}
{\cal H} &=&  h^{-1} \left(\tilde{\pi}_{ij} \tilde{\pi}^{ij} - \frac{1}{d-2} \tilde{\pi}^2\right) -(\mathcal{R}-2\Lambda) , \cr
{\cal H}^i &=& - h^{-1/2} D_j (2 \tilde{\pi}^{ij}).
\end{eqnarray}
In \eqref{Hz},
\begin{align}
G^{ijkl} &= h^{i(k}h^{l)j} - h^{ij}h^{kl},
\end{align}
$\kappa = 8\pi G$, $\partial\Sigma$ is the limit of constant $r = \sqrt{\delta_{ij}x^i x^j}$ submanifolds in $\Sigma$ as $r\rightarrow\infty$, and $\sigma_{ij}$ and $\hat{r}^i$ are the induced metric on and the unit normal (in $\Sigma$) to $\partial\Sigma$.

The boundary conditions are strong enough for (\ref{Hz}) to hold when general variations of the above boundary terms can be computed by varying only
$\Delta h_{ij}$ and $\Delta \tilde \pi^{ij}$; i.e., all other terms in the general variation are too small to contribute in the $r \rightarrow \infty$ limit.  Power counting shows that this is indeed implied by eqs. \eqref{BoundaryConditions}.

Now, naive power counting suggests that~\eqref{Hz} may diverge.  But since~\eqref{BoundaryConditions} ensures that the symplectic structure is finite, the equations of motion must conspire to prevent these divergences.  This fact is verified in appendix \ref{finiteQ}.  We define
\begin{equation}
\label{Qk=0}
Q_D := H(-\xi_D),  \ \ \ P^j := H(\xi_{P^j}), \ \ \ L^{jk} := H(\xi_{L^{jk}}).
\end{equation}
Note that in defining $Q_D$ we chose signs that would conventionally appear in the definition of an `energy,' while we defined $P^j$ and $L^{jk}$ with signs conventionally chosen in defining momenta.  Interestingly, this choice of signs makes $Q_D$ negative for the de Sitter-Schwarzschild solution in agreement with \cite{balasubramaniandS}.

We may also consider a general Hamiltonian $H(\partial_t)$ for $\partial_t$ defined by lapse and shift of the form \eqref{BoundaryConditions}.  Power counting and the boundary conditions~\eqref{BoundaryConditions} ensure that $H(\partial_t)$ is finite and that it has well-defined variations.  The boundary conditions~\eqref{BoundaryConditions} and~\eqref{lapseshiftcondition} ensure that it generates an evolution which preserves the boundary conditions on $h_{ij}, \tilde \pi^{ij}$ and thus, as noted in footnote \ref{symfoot}, that the symplectic structure is conserved.  It follows that the above charges are conserved as well.

In appendix~\ref{finiteQ} we explicitly show that that the conserved quantities~\eqref{Qk=0} are finite, and that with the boundary conditions~\eqref{BoundaryConditions} they take the following simple forms:
\begin{subequations} \label{Hk=0}
\begin{eqnarray}
Q_D  &=&  \frac{1}{\kappa}\int_{\partial\Sigma} \sqrt{\sigma} \frac{ \hat{r}_i   \Delta^\prime \pi^{ij} x_j}{\ell}   , \label{HDilatation} \\
P^j  &=&   \frac{1}{\kappa} \int_{\partial \Sigma} \sqrt{\sigma} \hat{r}_i  \Delta^\prime  \pi^{ij} , \label{Pi} \\
L^{jk}  &=&  \frac{1}{\kappa} \int_{\partial \Sigma} \sqrt{\sigma} \hat{r}_i 2\Delta^\prime \pi^{i[k} x^{j]}, \label{Li}
\end{eqnarray}
\end{subequations}
where
\begin{align}
\Delta^\prime \pi^{ij} = \pi^{ij} + \frac{d-2}{\ell} h^{ij}.
\end{align}

\subsection{Familiar Examples}
\label{FE}

We now consider some familiar spacetimes in order
to provide further intuition for the constructions above.  In particular, for $d \ge 4$ we find coordinates in which our phase space contains the de-Sitter Schwarzschild solution and we compute the relevant charges.  For $d=3$ we consider instead the spinning conical defect spacetimes, which describe gravity coupled to compactly supported matter fields.

In familiar static coordinates the $d \ge 4$ de Sitter-Schwarzschild solution takes the form
\begin{eqnarray}
ds^2 &=& - f(\rho) d\tau^2 + \frac{d\rho^2}{f(\rho)} + \rho^2 d\Omega^2 \label{dSschwarzMetric} \\
f(\rho) &=& 1-\frac{2GM}{\rho^{d-3}}-\frac{\rho^2}{\ell^2}.
\end{eqnarray}
In the region $\rho > \ell$ we introduce the coordinates $(t,\{x^i\})$ through the implicit expressions
\begin{subequations} \label{coordTrans}
\begin{eqnarray}
\tau &=& t + \int^\rho_\ell d\rho^\prime \frac{\sqrt{1-f(\rho^\prime)}}{f(\rho^\prime)} \\
\rho &=& e^{t/\ell}\sqrt{\sum_i (x^i)^2},
\end{eqnarray}
\end{subequations}
with the angular variables being related to $\{x^i\}$ in the usual way. After this change of coordinates \eqref{dSschwarzMetric} becomes
\begin{eqnarray}
\label{SdS2}
ds^2 = -dt^2 + e^{2t/\ell}\delta_{ij} \left(dx^i - \frac{ GM\ell x^i}{(e^{t/\ell}r)^{d-1}}  dt \right) \left(dx^j - \frac{ GM\ell x^j}{(e^{t/\ell} r)^{d-1}} dt \right) + \mathcal{O}(r^{-(2d-3)}),
\end{eqnarray}
for $r \gg \ell e^{-t/\ell}$. As a result, we find
\begin{subequations} \label{dsSchwarzhandpi}
\begin{eqnarray}
\Delta h_{ij} &=&  \mathcal{O}\left(r^{-(2d-3)}\right) \\
\Delta \pi^{ij} &=& \frac{e^{-2t/\ell}  GM\ell }{(e^{t/\ell}r)^{d-1}} \left(\delta^{ij} - (d-1)\frac{x^i x^j}{r^2}\right) + \mathcal{O}\left(r^{-(2d-3)}\right).
\end{eqnarray}
\end{subequations}
Comparison with \eqref{BoundaryConditions} shows that \eqref{SdS2} does indeed lie in the phase space $\CVPS$.

The linear and angular momenta for this solution vanish by symmetry.  Using~\eqref{HDilatation} to calculate the dilation charge yields
\begin{eqnarray}
\label{QDSdS}
Q_D &=& -\frac{(d-2)}{\kappa}GM \int_{\partial\Sigma} \sqrt{\gamma} d^{d-2}\theta \\
&=& -\frac{(d-2)\pi^{(d-3)/2}}{4\Gamma[(d-1)/2]}M,
\end{eqnarray}
where $\gamma$ is the determinant of the metric on the unit $S^{d-2}$ (so that the volume element on $S^{d-2}$ is $\sqrt{\gamma} d^{d-2}\theta$).  In particular, we find $Q_D = -M$ for $d=4$ and $Q_D = -(3\pi/4) M$ for $d=5$.  Up to a shift of the zero point of the energy for $d=5$ these results agree with the charges computed in \cite{balasubramaniandS} using a rather different approach (which did not involve constructing a phase space).  We will show in section \ref{compare} below that this agreement holds more generally\footnote{ Because the  counter-terms required by \cite{balasubramaniandS} proliferate in higher dimensions, section \ref{compare} considers only the cases $d=3,4,5$.  We expect similar results for higher dimensions. The charges of \cite{Strominger:2001pn} also differ by an overall sign. }.

We now turn to the  $d=3$ spinning conical defect solution \cite{Deser:1983nh} with defect angle $\theta_d$, which may be written in the form (see e.g. \cite{balasubramaniandS})
\begin{align}
ds^2 &= -f(\rho) d\tau^2 + \frac{d\rho^2}{f(\rho)} + \rho^2\left( \frac{8GMa}{2\rho^2} d\tau + d\phi\right)^2 \cr
f(\rho) &= 1 - 8GM - \frac{\rho^2}{\ell^2} + \frac{(8GMa)^2}{4\rho^2}, \label{ConicalDefect}
\end{align}
where the parameter $M=\theta_d/8\pi G$ is the mass that would be assigned to a conical defect in flat space with defect angle $\theta_d$~\cite{Deser:1983tn,Henneaux:1984ei}.\footnote{Our conventions are related to those of~\cite{balasubramaniandS} by $M\Rightarrow 1/8G - m$ and $aM \Rightarrow -J$.}  After changing to the coordinates $(t,r)$ defined by
\begin{align}
\tau &= t - \frac{\ell}{2}\log \left(\frac{e^{2t/\ell} r^2}{\ell^2} - 1\right) \\
\rho &= e^{t/\ell} r,
\end{align}
we find $\Delta h_{ij}, \Delta \pi^{ij}\sim \order{r}{-2}$ so that the transformed solution lies in $\CVPS$, though $h^{(2)}_{ij} \neq0$.  The non-vanishing results from \eqref{Hz} are $Q_D = -M$ and $L^{12}= aM$, in agreement with~\cite{balasubramaniandS} up to the expected shift in the zero point of $Q_D$ and in agreement with flat space results in the (here trivial) limit $\ell\rightarrow\infty$.

\subsection{Comparison with Brown-York methods at future infinity}
\label{compare}

Our discussion above closely followed the classic treatment of \cite{ReggeTeitelboim}.  In contrast, refs. \cite{Strominger:2001pn,balasubramaniandS} took a rather different approach to the construction of charges in de Sitter gravity.  They considered spacetimes for which the induced metric on {\it future} infinity $I^+$ (defined by a conformal compactification associated with a given foliation near $I^+$) agrees with some fixed metric $q_{ij}$.\footnote{In~\cite{balasubramaniandS} $q_{ij}$ is the flat metric, however one may clearly extend their results to allow non-flat $q_{ij}$ using a construction along the lines of~\cite{Henningson:1998gx}.  Such a construction would involve adding logarithmically divergent terms to~\eqref{BYCounterterm} for $d=3$ and $d=5$.  \label{logtermsFN}}
They then constructed an action $S$ for gravity subject to this Dirichlet-like boundary condition, choosing the boundary terms at future infinity so that variations of the action are well-defined.  By analogy with the Brown-York stress tensor of \cite{BrownYork} (and with the anti-de Sitter case \cite{Henningson:1998gx,balasubramanianStressTensor}), refs. \cite{Strominger:2001pn,balasubramaniandS} defined a de Sitter `stress-tensor' $\tau^{ij}$ on a $t=(\textrm{constant})$ slice (with the intention of eventually taking $t\rightarrow\infty$) through
\begin{eqnarray}
\label{BYST}
\tau^{ij} = \frac{2}{\sqrt{h}} \frac{\delta S}{\delta h_{ij}} =  \frac{\pi^{ij}}{\kappa}   + \tau^{ij}_{ct},
\end{eqnarray}
where the first term results from varying the Einstein-Hilbert action with a Gibbons-Hawking-like boundary term $1/\kappa \int_{I^+} \sqrt{q} K $ and $\tau^{ij}_{ct}$ is the result of varying so-called counter-terms added to the action.  Given a Killing field $\xi^i$ of $q_{ij}$ and any $d-2$ surface $B$ in $I^+$, \cite{Strominger:2001pn,balasubramaniandS} then define a charge
\begin{eqnarray}
Q_\xi(B) &=& \lim_{t\rightarrow\infty} \int_{B_t} \sqrt{\sigma} \left(\hat{b}_i \tau^{ij}  \xi_j\right), \label{BYCharge}
\end{eqnarray}
where $\hat{b}^i$ and $\sqrt{\sigma}$ are the unit normal to and induced volume element on $B_t$ and $B_t$ is a $d-2$ surface on a constant $t$ slice that approaches $B$ as $t\rightarrow\infty$.  Since $\xi^i$ is a Killing field of $q_{ij}$,  the charge is in fact independent of $B$.  For even $d$ one can show that $\tau^{ij}$ is traceless so that this $B$-independence also holds when this definition of charge is extended to conformal Killing fields $\xi^i$.

Because $q_{ij}$ is fixed, these boundary conditions force the symplectic flux through $I^+$ to vanish.  So long as all other boundary conditions (e.g., at $r = \infty$) enforce conservation of symplectic flux, it follows immediately that this flux also vanishes on any Cauchy surface.  As a result, the class of spacetimes for which $S$ is a valid variational principle does not form a phase space (though see \cite{Anninos:2011jp} for further discussion).  On the other hand, as shown in \cite{StromingerASG}, the charges \eqref{BYCharge} agree with a natural construction that does not require the condition of fixed $q_{ij}$, but which is instead given by the covariant phase space prescription used by Wald and Zoupas \cite{Wald:1999wa} to define charges for the Bondi-Metzner-Sachs group in asymptotically flat space \cite{Bondi:1962px,Sachs:1962wk,Sachs:1962zza,Penrose:1962ij}.\footnote{This makes it clear that the charges of \cite{JagerPhD} also agree. These charges were constructed using Noether charge methods with fixed $q_{ij}$.}  In this context, one takes $\xi^i$ above to be an arbitrary vector field on $I^+$ (and one generally expects $Q_\xi(B)$ to depend on $B$).  Though it is not immediately clear in what sense such charges generate symmetries, this fact nevertheless suggests that the charges \eqref{BYCharge} are of interest even when $q^{ij}$ is not fixed.  This is also suggested by the formal analogy with anti-de Sitter space.

In any case, we saw earlier that when $B$ is taken to be $i^0$  (as defined by figure \ref{conformalDiagram}) for $k=0$, $q_{ij}$ is taken to be the metric on the surface $t=\infty$, and  $\xi^a$ is a generator of asymptotic symmetries for $\CVPS$, the charges \eqref{BYCharge} coincide with ours (up to a shift of the zero of energy) for the particular cases of $d=4,5$ de Sitter-Schwarzschild  and the $d=3$ spinning conical defect in appropriate coordinates.  It might seem natural to suppose that this equivalence extends to all of $\CVPS$.  Such a correspondence is plausible since near $i^0$ the boundary conditions for $\CVPS$ require the spacetime to approach exact de Sitter space and the induced metric on $I^+$ becomes approximately fixed.  We show below that in $d=3,4,5$ for generators of our asymptotic symmetries the charges \eqref{BYCharge} actually yield precisely \eqref{HDilatation}, \eqref{Pi}, \eqref{Li} (up to a possible shift of the zero points).  They thus agree with our charges.

We wish to take $B$ to lie at $i^0$, which we will think of as the boundary $\partial \Sigma_\infty$ of the surface $\Sigma_\infty$ on which $t =\infty$.  The unit normal to $\partial \Sigma_\infty$ in $\Sigma_\infty$ is thus $\hat b^i = \hat r^i$.  We use the fact that, since any asymptotic symmetry $\xi^a$ preserves $I^+$, it defines a vector field $\xi^i_{I^+}$ on $I^+$.  In fact, this $\xi^i_{I^+}$ is just the $t \rightarrow \infty$ limit of the part $\vec \xi^i$ of $\xi^a$ tangent to $\Sigma$ as defined by \eqref{parts}.  We therefore use the notation $\vec \xi^i$ for this vector field below.  It will further be useful to decompose $\vec \xi^i$ into parts normal and tangent to $\partial \Sigma_{\infty}$ according to
\begin{equation}
\label{2ndparts}
\vec \xi^i = \vec \xi_\perp \hat r^i  + \vec {\vec \xi}^i.
\end{equation}

To compute the counter-term charges we recall from \cite{Strominger:2001pn,balasubramaniandS} that for $d=3, 4, 5$, (see footnote~\ref{logtermsFN})
\begin{eqnarray} \label{BYCounterterm}
\tau^{ij}_{CT} = \frac{1}{\kappa}\left(\frac{(d-2)h^{ij}}{\ell} + \frac{\ell}{d-3} \mathcal{G}^{ij} \right), 
\end{eqnarray}
where $\mathcal{G}^{ij}$ is the Einstein tensor of $\Sigma_\infty$ and the $\mathcal{G}^{ij}$ term does not appear for $d=3$.  It is then clear that first term in \eqref{BYCounterterm} combines nicely with the explicit $\pi^{ij}$ term in \eqref{BYST} to give a term involving $\Delta^\prime \pi^{ij}$; i.e., this piece of the counterterm cancels the contribution from the pure $\dS$ background.  For $d=3$ we then see that
\eqref{BYCharge} precisely reproduces ~\eqref{HDilatation},~\eqref{Pi}, and~\eqref{Li}, and the same would be true for $d=4,5$ if the $\mathcal{G}^{ij}$ term in \eqref{BYCounterterm} can be ignored.

This is in fact the case, as we now show for $d=4,5$ that the $\mathcal{G}^{ij}$  term in \eqref{BYCounterterm} is independent of $(\Delta h_{ij},\Delta \pi^{ij})$ and thus yields at most an irrelevant shift of the zero point of the charge\footnote{By symmetry, such a shift is allowed only for the $Q_D$.}.  To do so, recall that
$\hat r_i \mathcal{G}^{ij}$ can be expressed in terms of the Ricci scalar $\mathfrak{R}$ and the extrinsic curvature  of $\partial \Sigma_\infty$ through the Gauss-Codacci equations.  We will work with the pull-back $\theta_{ij}$ of this extrinsic curvature to $\Sigma_\infty$.

In particular, the contribution involving $\vec \xi_\perp$ is related to the `radial Hamiltonian constraint'
\begin{eqnarray}
\label{rH}
\mathcal{G}_{ij}\hat{r}^i \hat{r}^j = -\frac{1}{2}(\mathfrak{R} - \theta^2 + \theta_{ij}\theta^{ij}).
\end{eqnarray}
After expanding $h_{ij} = \bar h_{ij} + \Delta h_{ij}$,
power counting shows that only the (constant) background term and terms linear in $\Delta h_{ij}$ can contribute in the $r\rightarrow \infty$ limit.  A bit of calculation (given in appendix \ref{details}) then shows
\begin{eqnarray}
\label{pp}
\frac{\hat r^i \ell \mathcal{G}_{ij} \vec{\xi}_\perp \hat r^j}{d-3}  = -\frac{\ell \vec{\xi}_\perp}{2 r}   \hat{r}^m \bar{h}^{jk} \left(D_j \Delta h_{km} -  D_m \Delta h_{jk}\right) + (\text{constant}) + \dots,
\end{eqnarray}
where $\dots$ represents both higher order terms (that do not contribute as $r \rightarrow \infty$) and total divergences on $\partial \Sigma_\infty$.  Power counting again then shows that the non-trivial term in \eqref{pp} vanishes by our boundary conditions.

We now turn to the counter-term contribution involving $\vec {\vec \xi}^i$, which is a combination of the `radial momentum constraints':
 \begin{equation}
 \label{mc}
 \hat r_i \mathcal{G}^{ij} \vec {\vec \xi}_j = \mathcal D_i \left( \theta^{ij} - \theta \sigma^{ij} \right) \vec {\vec \xi}_j,
 \end{equation}
where $\cal D$ is the derivative operator associated with $\sigma_{ij}$.  We now treat the various asymptotic symmetries separately:
This term vanishes explicitly for dilations as they have $\vec {\vec \xi}^i =0$.  For rotations, the symmetry of $\theta^{ij}$ and the fact that  $\vec {\vec \xi}^i$ is a Killing field of $\partial \Sigma_\infty$ allow us to bring the factor of $\vec {\vec \xi}^i =0$ inside the parentheses and write \eqref{mc} as a total divergence on $\partial \Sigma$. Thus its integral over $\partial \Sigma$ (a closed manifold) must vanish.   Finally for translations we use the fact that $\vec {\vec \xi}^i$ is a conformal Killing vector of $\partial\Sigma$ to write
\begin{align}
\hat r_i \mathcal{G}^{ij} \vec {\vec \xi}_j = \frac{(d-3)\theta}{r}  \vec\xi_\perp.
\end{align}
The leading order contribution to this term vanishes upon integration due to the fact that $\vec\xi_\perp$ is odd.  The remaining terms vanish by power counting.  It follows that \eqref{mc} makes no contribution to the total charge and we see that, as previously claimed, the charges \eqref{BYCharge} are given up to a possible shift of the zero-point by \eqref{HDilatation}, \eqref{Pi}, \eqref{Li}.

\section{The $k=-1$ phase space}
\label{k=-1}

\begin{figure}
\includegraphics[width=4.5in]{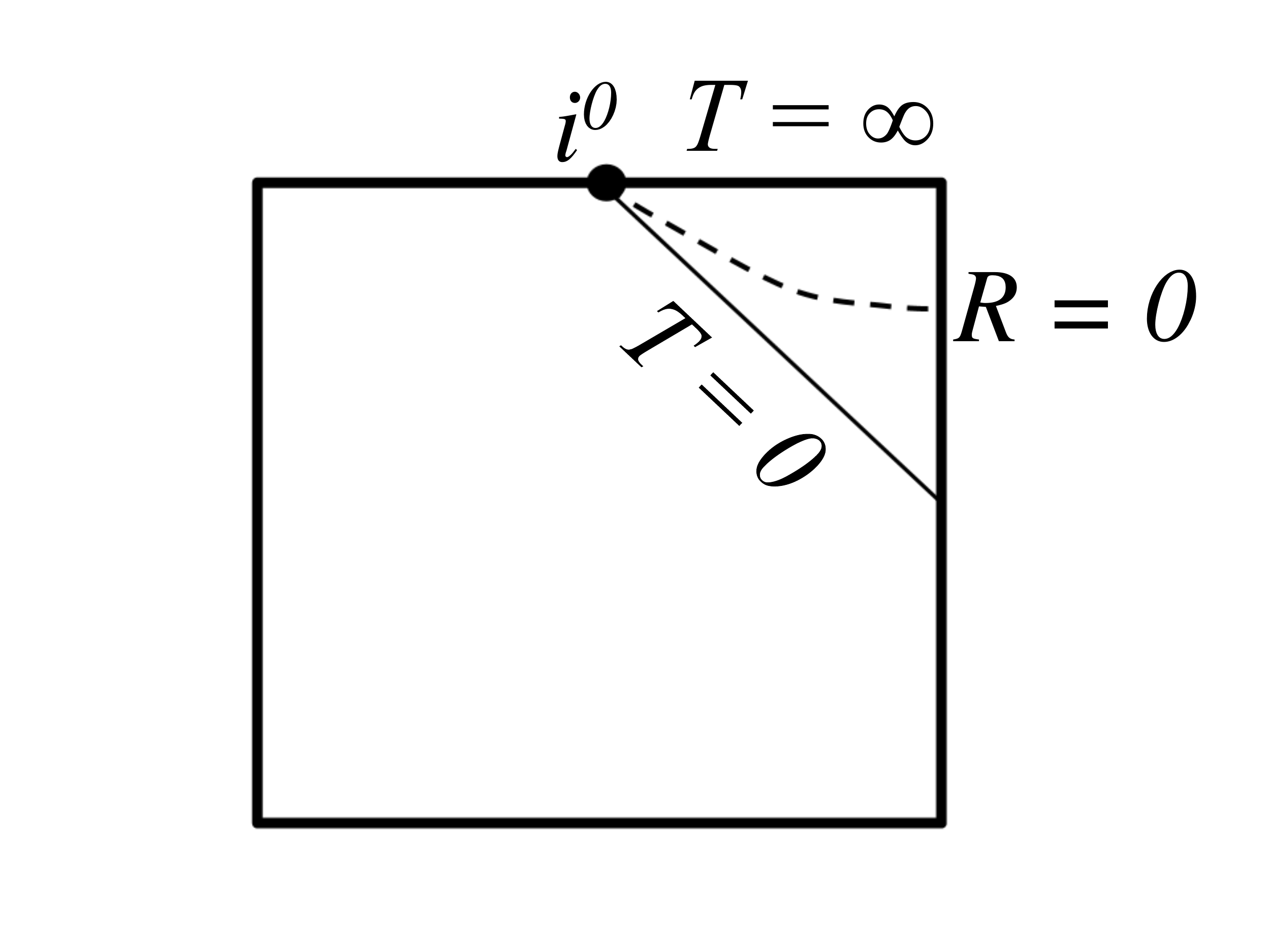}
\caption{A conformal diagram of dS${}_d$.  The region above the diagonal line is the expanding hyperbolic patch.  The $S^{d-2}$ labeled $i^0$ is the spatial infinity at which our boundary conditions are applied.  The dashed line shows a representative $(T= {\rm constant})$ slice.}
\label{conformalDiagramHyp}
\end{figure}

We now construct two phase spaces of $d \ge 3$  spacetimes which asymptotically approach the hyperbolic ($k=-1$) patch of de Sitter space (see figure ~\ref{conformalDiagramHyp}) with metric
\begin{eqnarray}
ds^2 = -dT^2 + \sinh^2(T/\ell)\left(\delta_{ij} - \frac{1}{1+\ell^2/R^2}\frac{X_i X_j}{R^2}\right) dX^i dX^j, \label{dsHypVacuum}
\end{eqnarray}
where $X_i := \delta_{ij}X^j$. The Killing vectors then take the simple form
\begin{eqnarray}
\label{k=-1KVFs}
{\textrm {`Translations'}:} & \xi_{P^i}^a = \sqrt{R^2+\ell^2} (\partial_i)^a, \cr
{\rm Rotations:} & \xi_{L^{ij}}^a = (2X_{[i}\partial_{j]})^a.
\end{eqnarray}
These are precisely the Killing fields of $H^{d-1}$ and so generate the Lorentz group $SO(d-1,1)$.  One might thus equally well refer to the hyperbolic `translations' as boosts.

The phase space $\CVPSh$ will be constructed in direct analogy to our treatment for $k=0$ in section \ref{k=0}.  We again consider globally hyperbolic solutions to Einstein's equations in $d\ge 3$ spacetime dimensions with positive cosmological constant and topology $\mathbb{R}^d$.  Introducing a foliation as before with coordinates $(T,\{X^i\})$ and a metric of the form \eqref{dSmetric}, we define $\CVPSh$ to contain spacetimes with induced metric $h_{ij}$ canonical momentum ${\pi}^{ij}$, lapse $N$, and shift $N^i$ on a $(T= {\rm constant})$ slice $\Sigma$ which satisfy
\begin{align}
\Delta h_{ij} &= h^{(d-2)}_{ij} + \order{h^{(d-2)}_{ij} R}{-1} \cr
\Delta {\pi}^{ij} &= \pi^{ij}_{(d-2)} + \order{\pi_{(d-2)}^{ij} R}{-1} \cr
N  &= 1+ N^{(d-2)} + \order{R}{-(d-1)} \cr
N^i &= N^i_{(d-3)}+\order{N^i_{(d-3)}R}{-1}, \label{BoundaryConditionsh}
\end{align}
for large $R$ with $R=\sqrt{\delta_{ij}X^i X^j}$.    The required falloff of $h^{(d-2)}_{ij}$, $\pi^{ij}_{(d-2)}$ and $N^i_{(d-3)}$ are most clearly expressed in spherical coordinates,\begin{subequations}
\begin{align}
\label{hdm2}
h_{RR}^{(d-2)} &=  \frac{(\textrm{Function of $\Omega$})}{R^{2+(d-2)}} ,  &  h_{R\theta}^{(d-2)} &=  \frac{(\textrm{Function of $\Omega$})}{R^{(d-2)}} ,  &   h_{\phi\theta}^{(d-2)} &=  \frac{(\textrm{Function of $\Omega$})}{R^{-2+(d-2)}} \\
\pi^{RR}_{(d-2)} &= \frac{(\textrm{Function of $\Omega$})}{R^{-2+(d-2)}} ,  &         \pi^{R\theta}_{(d-2)} &= \frac{(\textrm{Function of $\Omega$})}{R^{(d-2)}} ,  &  \pi^{\phi\theta}_{(d-2)} &= \frac{(\textrm{Function of $\Omega$})}{R^{2+(d-2)}} \\
N^{(d-2)} &= \frac{(\textrm{Function of $\Omega$})}{R^{(d-2)}} , &  N_{(d-3)}^R &= \frac{(\textrm{Function of $\Omega$})}{R^{(d-3)}} , & N_{(d-3)}^\theta &= \frac{(\textrm{Function of $\Omega$})}{R^{2+(d-3)}}
\end{align}
\end{subequations}
with $\theta$ and $\phi$ standing in for any angular coordinates.  We define
\begin{subequations} \label{DeltahpiHyp}
\begin{eqnarray}
\Delta h_{ij} &=& h_{ij} - \sinh^2\left(T/\ell\right)\omega_{ij} \\
\Delta {\pi}^{ij} &=& {\pi}^{ij} + \frac{(d-2) \coth(T/\ell) \sinh^{-2}(T/\ell)}{\ell}{\omega}^{ij}.
\end{eqnarray}
\end{subequations}
  Here we have introduced the metric $\omega_{ij}$ on the unit $H^{d-1}$:
\begin{eqnarray}
\omega_{ij} = \delta_{ij} - \frac{1}{1+\ell^2/R^2}\frac{X_i X_j}{R^2}.
\end{eqnarray}
The definitions \eqref{DeltahpiHyp} ensure that $\Delta h_{ij}=0=\Delta{\pi}^{ij}$ for $\dSh$.  The boundary conditions~\eqref{BoundaryConditionsh} are sufficient to ensure that, in spherical coordinates,
\begin{align}
\Delta\tilde\pi^{ij} = \order{\Delta\pi_{(d-2)}^{ij}R}{d-3},
\end{align}
which makes the symplectic structure \eqref{SymplecticStructure} finite.  We will show below that it is also conserved.  Thus $\CVPSh$ is a well-defined phase space.

\subsection{Asymptotic Symmetries}

As in the $\dS$ case we know that any element of our ASG must map~\eqref{dsHypVacuum} onto a spacetime satisfying~\eqref{BoundaryConditionsh}.  This means that the associated vector field must satisfy $h^a_i h^b_j \bar \nabla_{(a} \xi_{b)}~=~\mathcal{O}\left(R^{-(d-2)}\right)$, or
\begin{align}
D_{(i}\vec\xi_{j)} &=  \partial_{(i}\vec\xi_{j)} + \frac{R \vec\xi_\perp}{\ell^2} \omega_{ij} =  \frac{\sinh(T/\ell)\cosh(T/\ell)\xi_\perp}{\ell} \omega_{ij}, \label{ckvfHyp}
\end{align}
up to terms which vanish at infinity, where $D_i$ is the covariant derivative on $\Sigma$.  For $d\ge 4$ we project this equation onto a $R~= ~(\text{constant})$ submanifold which gives
\begin{align}
\mathcal D_{(\underline i}\vec{\vec\xi}_{\underline{j})} &= \left(\frac{\sinh(T/\ell)\cosh(T/\ell)\xi_\perp}{\ell}-\frac{(1-R^2/\ell^2)\vec\xi_\perp}{R} \right) \sigma_{\underline{ij}} ,  \label{ckvfHypP}
\end{align}
where $\mathcal D_i$ is the covariant derivative on the $R= (\text{constant})$ subsurface of $\Sigma$.  Since we consider here the metric \eqref{dsHypVacuum}, eqn. \eqref{ckvfHypP} is the conformal Killing equation on the unit $S^{d-2}$.   The sphere is conformally flat, so the solutions to this equation are the generators of the $d-2$ dimensional Euclidean conformal group.  For $d\ge 5$ this group is $SO(d-1,1)$ which is isomorphic to the group generated by \eqref{k=-1KVFs}.  For $d=4$, \eqref{ckvfHypP} has an infinite number of solutions, however we are only interested in those solutions which are globally well defined on the sphere.  These solutions form the subgroup $PSL(2,\mathbb{C})~\cong~SO(3,1)$, which is again isomorphic to the group generated by \eqref{k=-1KVFs}.  We conclude that our $d \ge 4$ ASG can only contain symmetries which asymptotically approach the isometries \eqref{k=-1KVFs}.  The case $d=3$ will be addressed below.

As before,~\eqref{LieDhpi} shows that our phase space is closed under the isometries of $\dSh$ (noting that $\vec\xi\sim R$ and $\xi_\perp = 0$).  So we have shown that the asymptotic symmetry group of $\CVPSh$ is isomorphic to the isometries of $\dSh$ for $d\ge 4$. Using the same technique as in section \ref{CC} (and appendix~\ref{finiteQ}), we obtain conserved charges for the asymptotic symmetries of $\CVPSh$ and $d\ge4$, 
\begin{eqnarray} \label{Qk=-1}
P^j &\equiv& H(\xi^a_{P^j}) = \frac{1}{\kappa}\int_{\partial\Sigma} \sqrt{\sigma}  R \hat{R}_i\Delta^\prime \pi^{ik} h_{kj}   \label{Phyp} \cr
L^{jk} &\equiv& H(\xi^a_{L^{jk}}) = \frac{1}{\kappa}\int_{\partial\Sigma} \sqrt{\sigma} 2  \hat{R}_i \Delta^\prime \pi^{im} \bar h_{ml} \delta^{l[k} X^{j]}, \label{Lhyp}
\end{eqnarray}
with
\begin{align}
\Delta^\prime \pi^{ij} = \pi^{ij} + \frac{(d-2) \coth(T/\ell) }{\ell}  h^{ij}.
\end{align}
The charges~\eqref{Qk=-1} are finite by the boundary conditions~\eqref{BoundaryConditionsh}.  As for $k=0$ one finds that  $H(\partial_T)$ is finite, has well-defined variations, and generates an evolution that preserves the boundary conditions~\eqref{BoundaryConditionsh} on $h_{ij},\tilde \pi^{ij}$. (For $k=-1$ there is no need to introduce an analogue of \eqref{lapseshiftcondition}.)  So we again conclude that the symplectic structure and the above charges are conserved on $\CVPSh$.

The case $d=3$ is special due to the infinite-dimensional conformal group in two dimensions.  Since the hyperbolic plane is conformally flat, the solutions of~\eqref{ckvfHyp} define two commuting Virasoro algebras formally associated with charges $L_n$ and $\bar L_n$ for $n \in {\mathbb Z}$ satisfying $L_n^* = \bar L_{-n}$ where $*$ denotes complex conjugation and $n$ labels the angular momentum quantum number.
 The details of the vector fields are given in appendix \ref{d=3ASG}.
 The unfamiliar reality condition is due to the fact that the symmetries of the Lorentz-signature theory generate the 2d Euclidean-signature conformal group and was previous discussed in \cite{Fjelstad:2002wf,Balasubramanian:2002zh,Ouyang:2011fs}.    With our conventions, the angular momentum (called $L_{ij}$ above in higher dimensions) is $J_0$ where $J_n = (L_n + \bar L_n)$.  It is also useful to introduce $K_n = (L_n - \bar L_n)/i\ell$.  We will see below that $K_0$ captures energy-like information about solutions.

As noted in section \ref{k=0}, expression \eqref{Hz} is valid only when second order terms in $\Delta h_{ij}, \Delta \pi^{ij}$ do not contribute to the variations. For the $J_n$ charges, this condition holds on all of $\CVPSh$.  For the $K_n$ charges, it holds only when we use the gauge freedom to set $h^{(1)}_{ij} =0$. This is always possible for $d=3$ since $h_{ij}$ has only 3 independent components.  With this understanding for $K_n$, we find
\begin{align}
 J_n &\equiv H(\xi_{J_n}) = \frac{1}{\kappa}\int_{\partial\Sigma} \sqrt{\sigma} 2 \hat{R}_i\Delta^\prime \pi^{im} \bar h_{ml} \delta^{l[k} X^{j]} e^{i n \theta} \cr
K_n &\equiv H(-\xi_{K_n}) =  \frac{1}{\kappa}\int_{\partial\Sigma} \sqrt{\sigma} \hat{R}_i \Delta^\prime \pi^{ik} \bar h_{kj} \frac{R^2\coth(T/\ell)}{\ell^2} (\partial_R)^j e^{in\theta} \cr
& \quad + \frac{1}{2\kappa} \int_{\partial\Sigma} \sqrt{\sigma}  \hat{R}_l G^{ijkl} \left(-\frac{R}{\ell} D_k \Delta h_{ij}+\frac{\omega_k^R}{\ell} \Delta h_{ij} \right)e^{in\theta}. \label{V2}
\end{align}
In choosing sign conventions we have treated $J_n$ as a momentum and $K_n$ as an energy.  Since appendix \ref{KnConservation} shows that the expression for $K_n$ above is invariant under gauge transformations that preserve $h^{(1)}_{ij} =0$, we may use \eqref{V2} to define $K_n$ as a gauge invariant charge on all of $\Gamma(dS_{k=-1})$.

From the Poincar\'e disk description of the 2d hyperbolic plane, one readily sees that diffeomorphisms associated with $J_n$ preserve the boundary while those associated with $K_n$ do not.  A careful study of our boundary conditions \eqref{BoundaryConditionsh} similarly shows that the phase space $\CVPSh$ is invariant only under the $J_n$. As a result, the asymptotic symmetry group of either $\CVPSh$ or $\CVPSh_{gf}$ is given by a single Virasoro algebra
\begin{align}
[J_{n}, J_{_m} ] =  (n-m) J_{n+m},
\end{align}
where $[A,B]$ denotes the commutator of the corresponding quantum mechanical charges (i.e., we have inserted an extra factor of $i$ relative to the classical Poisson Bracket). The central charge vanishes due to the symmetry $\theta \rightarrow - \theta$, which reflects the lack of a gravitational Chern-Simons term.  Adding such a term to the action should lead to non-vanishing central charge.

Simple power counting shows that $J_n$ is finite on $\CVPSh$, and also that  $H(\partial_T)$ is finite, has well-defined variations, and generates an evolution that preserves the boundary conditions~\eqref{BoundaryConditionsh} on $h_{ij},\tilde \pi^{ij}$. It follows that the symplectic structure and the $J_n$ charges are conserved on $\CVPSh$, and thus on $\CVPSh_{gf}$ as well.  While the $K_n$ do not generate asymptotic symmetries, it turns out that expresion~\eqref{V2} is nevertheless finite, gauge invariant, and time independent on $\CVPSh$.  The details of this argument are given in Appendix~\ref{KnConservation}.

We take these observations as motivation to consider further the full 2d conformal algebra.  The associated central charges can be computed as in \cite{BrownHenneauxCentralCharge} and turn out to be non-zero:
\begin{align}
\label{ccharge}
[L_n, L_m] &= (n-m) L_{n+m} + i \frac{1}{12} \left(\frac{3\ell}{2 G}\right)  n(n^2-1) \delta_{n,-m} \\
[\bar L_n, \bar L_m] &= (n-m) \bar L_{n+m}  - i \frac{1}{12} \left(\frac{3\ell}{2 G}\right)  n(n^2-1) \delta_{n,-m} \\
[L_n, \bar L_m] &= 0 ,
\end{align}
so that the left- and right-moving central charges are imaginary complex conjugates
in agreement with \cite{Balasubramanian:2002zh,Maldacena:2002vr,Ouyang:2011fs}.

It would be interesting to understand the unitary representations of (\ref{ccharge}) under the appropriate reality conditions.  This question was briefly investigated in \cite{Balasubramanian:2002zh}.  However, our analysis suggests that there is an additional subtlety:  Because the flows generated by $K_n$ are not complete on our classical phase space, ``real'' elements of the algebra they generate  (e.g., $K_0$) are unlikely to be self-adjoint on the quantum Hilbert space.   Indeed, it is natural to expect behavior resembling that of $-i\frac{d}{dx}$ on the half-line, which admits complex eigenvalues.  This in principle allows representations more general than those considered in  \cite{Balasubramanian:2002zh}, though we will not pursue the details here.

For later use, we note that imposing the additional gauge condition 
\begin{equation}
\label{ggh}
\Delta h_{ij} = h_{ij}^{(3)} + \order{h_{ij}^{(3)} R}{-\epsilon}
\end{equation}
further simplifies \eqref{V2} and yields:
\begin{align}
\label{simple}
J_n &=  \frac{1}{\kappa}\int_{\partial\Sigma} \sqrt{\sigma} \hat R_i 2  \Delta^\prime \pi^{im} h_{ml} \delta^{l[k} X^{j]} e^{i n \theta} \cr
&=  \frac{1}{\kappa}\int_{\partial\Sigma}  \pi_{(2)}^{R\theta} R^2 \ell \sinh^4(T/\ell) e^{i n \theta} \cr
K_n &= \frac{1}{\kappa} \int_{\partial\Sigma} \sqrt{\sigma} \hat{R}_i \Delta^\prime \pi^{ik} h_{kj}  \frac{R^2\coth(T/\ell)}{\ell^2} (\partial_R)^j e^{in\theta} \cr
&=  \frac{1}{\kappa} \int_{\partial\Sigma}  \pi_{(2)}^{RR}  \sinh^3(T/\ell) \cosh(T/\ell) e^{in\theta}.
\end{align}

\subsection{Familiar Examples}
\label{exk=-1}

We now study two classes of familiar solutions -- the $d=4$ Kerr-de Sitter solution and the $d=3$ spinning conical defect.  We find coordinates for which each solution lies in the phase space $\CVPSh$ and compute the appropriate charges.  We consider the Kerr case (as opposed to just Schwarzschild) since spherical symmetry would force all $d=4$ charges to vanish.

Our first task is to transform the standard $d=4$ Kerr-de Sitter metric~\cite{carterKerrdS}
\begin{eqnarray}
ds^2 &=& -\frac{\Delta-\Sigma a^2\sin^2(\psi)}{\Omega} d\tau^2 + a\sin^2(\psi)\left(\frac{2GM\rho}{\Omega}+\frac{\rho^2+a^2}{\ell^2}\right) (dt d\gamma+d\gamma dt) + \frac{\Omega}{\Delta} d\rho^2 \cr
 &+& \frac{\Omega}{\Sigma} d\psi^2 + \sin^2(\psi)\left(\frac{2GM\rho a \sin^2(\psi)}{\Omega}+\Delta+2GM\rho\right)d\gamma^2 \label{KdS} \cr
\Delta &=& (\rho^2 + a^2)\left(1-\frac{\rho^2}{\ell^2}\right)-2GM\rho \cr
\Omega &=& \rho^2+a^2\cos^2(\psi) \cr
\Sigma &=& 1+\frac{a^2}{\ell^2}\cos^2(\psi), \label{KerrMetric}
\end{eqnarray}
into coordinates for which it satisfies the boundary conditions \eqref{BoundaryConditionsh}.

We proceed by introducing coordinates $(s,\theta,\phi)$ through the expressions (c.f. appendix B of~\cite{HenneauxTeitelboimAsympAdS})
\begin{align}
s\cos(\theta) &= \rho\cos(\psi) \cr
\left(1+\frac{a^2}{\ell^2}\right)s^2 &= \rho^2+a^2\sin^2(\psi)+\frac{a^2}{\ell^2} \rho^2 \cos^2(\psi) \cr
\phi &= \left(1+\frac{a^2}{\ell^2}\right) \gamma + \frac{a}{\ell^2} \tau. \label{KerrcoordTrans}
\end{align}
In $(\tau,s, \theta, \phi)$ coordinates the metric  \eqref{KdS} approaches exact de Sitter space in static coordinates as $M\rightarrow 0$.  We then introduce further coordinates $T,R$ through
\begin{align}
\tau &= \ell \log\left(\frac{\cosh(T/\ell)+\sinh(T/\ell)\sqrt{1+R^2/\ell^2}} {|\sinh^2(T/\ell)R^2/\ell^2-1|^{1/2}}\right) \cr
s &= \sinh(T/\ell) R. \label{TRcoord}
\end{align}
By transforming \eqref{KdS} to $(T,R,\theta,\phi)$ coordinates we obtain a metric which approaches \eqref{dsHypVacuum} when $M \rightarrow 0$.  The explicit form of the metric is unenlightening but yields $\Delta h_{ij},\Delta \pi_{ij}$
which satisfies \eqref{BoundaryConditionsh}.  A similar calculation using the de Sitter-Schwarzschild solution in $d \ge 4$ yields fields with $h_{ij}^{(d-2)}=0=\pi^{ij}_{(d-2)}$ and which again satisfy \eqref{BoundaryConditionsh}.   Thus we see that our phase spaces are non-trivial for $d \ge 4$.  Note that in $d=4$ the leading order terms in $(\Delta h_{ij},\Delta \pi^{ij}$) for rotating black holes vanish as $a\rightarrow 0$.  We expect the same to be true in higher dimensions, though with the rotating solutions still satisfying \eqref{BoundaryConditionsh}.

Returning to the Kerr-de Sitter solution, symmetry implies that the only non vanishing charge is $L^{12}$.  From \eqref{Lhyp} we find
\begin{eqnarray}
L^{12} = \frac{aM}{(1+a^2/\ell^2)^2}. \label{KerrJ}
\end{eqnarray}
This differs from the analogous AdS result~\cite{HenneauxTeitelboimAsympAdS} only by the expected replacement $\ell\rightarrow i\ell$ and agrees with the analogous flat space result when $\ell\rightarrow\infty$.\footnote{Note that \cite{HenneauxTeitelboimAsympAdS} used a different sign convention for $a$.}

Finally, for $d=3$ we once again consider the conical defect solution~\eqref{ConicalDefect} of~\cite{BrownHenneauxCentralCharge,balasubramaniandS}.  After transforming from static to $k=-1$ coordinates (through a transformation resembling~\eqref{TRcoord}) we find $\Delta h_{ij}, \Delta \pi^{ij} $ satisfy~\eqref{BoundaryConditionsh} with $h_{ij}^{(1)}=0$.  Thus the  transformed solution is in $\CVPSh$.  We find $J_0 = aM$ and $K_0 = -M$.  Note that this agrees with $Q_D$ as computed for the $k=0$ version of the conical defect in section \ref{FE}.  As will be clear after we show equivalence to the counter-term charges in section \ref{comparehyp} below,  this is due to the fact that $K_0$ and $Q_D$ correspond to the same element of the Euclidean conformal group on $I^+$.

\subsection{Asymptotically dS${}_{k=-1}$ wormhole spacetimes}
\label{wormholes}

We now turn to some more novel spacetimes asymptotic to dS${}_{k=-1}$.  We consider pure $\Lambda > 0$ Einstein-Hilbert gravity, for which all solutions are quotients of dS${}_3$.  As noted in \cite{Balasubramanian:2002zh}, quotients generated by a single group element fall into two classes (up to congujation): The first leads to the conical defects discussed above.  For the 2nd class, the generator of the quotient group can be described simply in terms of the 3+1 Minkowski space $M^{3,1}$ into which dS${}_3$ is naturally embedded (as the set of points of proper distance $\ell$ from the origin).  This generator then consists of a simultaneous boost (say, along the $z$-axis) and a commuting rotation (in the $xy$ plane).  This class of quotients was not investigated in \cite{Balasubramanian:2002zh}, essentially because the resulting spacetimes are not asymptotically $\dSh$.  Indeed, from the point of view of the $k=0$ patch the quotient spacetime is naturally interpreted as a cosmological solution in which space is a cylinder $(S^1 \times \mathbb{R})$ at each moment of time.

However, in appropriate coordinates this 2nd class of quotients also defines spacetimes asymptotic to the $k=-1$ patch and which in fact lie in $\CVPSh$.  In this sense our quotient spacetimes may be thought of as $\Lambda > 0$ analogues of BTZ black holes \cite{Banados:1992wn,Banados:1992gq}.  That the quotient lies in $\CVPSh$ is easy to see when the quotient generator is a pure boost (i.e., where the commuting rotation is set to zero) in which case the quotient group preserves the appropriate $k=-1$ patch.   Indeed,
recall that for non-spinning BTZ black holes the associated quotient on AdS${}_3$ acts separately on each slice of constant global time, and that such surfaces are two-dimensional hyperbolic space $H^2$.  Here the $T=(\text{constant})$ slices of \eqref{dsHypVacuum} are also $H^2$ and we apply the analogous quotient.
This amounts to defining new coordinates $(R,\theta)$ through
\begin{equation}
X = \frac{2\pi R}{\theta_0}, \ \ \
Y = \sinh\left(\frac{\theta_0}{2\pi} \theta\right) \sqrt{\left(\frac{2\pi R}{\theta_0}\right)^2 +\ell^2},
\end{equation}
and taking $\theta$ to be periodic with period $2\pi$. The $T= (\text{constant})$ slices are then topologically $S^1\times \mathbb{R}$ and the the metric is
\begin{align}
\label{nsw}
g_{ab} = -dT^2 + \sinh^2(T/\ell) \left(\frac{\ell^2  dR^2}{R^2 + (\theta_0\ell/2\pi)^2} +  \left(R^2 + (\theta_0\ell/2\pi)^2\right) d\theta^2\right).
\end{align}
It is evident that \eqref{nsw} satisfies \eqref{BoundaryConditionsh} (in fact, with $h^{(1)}_{ij}=0$) and thus lies in $\CVPSh$.  We refer to \eqref{nsw} as a `wormhole' since on a given constant $T$ surface the $\theta$ circle has a minimum size $ \theta_0 \ell \sinh(T/\ell)$ at $R=0$.  These solutions have $J_0 =0$ and  $K_0 = - \left(1+(\theta_0/2\pi)^2\right)/8G$.  Since $K_0$ is an energy-like charge that vanishes for $dS_{k=-1}$, we find a mass gap analogous to that between AdS${}_3$ and the BTZ black holes.

When the commuting rotation is non-zero, the quotient group does not preserve the $k=-1$ patch of exact dS${}_3$.  Yet it appears that the quotient can nevertheless be considered to lie in $\CVPSh$.  Indeed, assuming a single rotational symmetry it is straightforward to solve the constraints \eqref{constraints} to find initial data for wormholes with angular momentum lying in $\dSh$.  For data asymptotic to a $T= (\text{constant})$ surface we find
\begin{align} \label{sw}
h_{ij} &= \sinh^2(T/\ell) \left(
\begin{array}[c]{cc}
\dfrac{\ell^2}{R^2+(\theta_0\ell/2\pi)^2}  & 0 \\
0 & R^2 +(\theta_0 \ell/2\pi)^2
\end{array} \right) \cr
\pi^{ij}  &= -\frac{\coth(T/\ell)}{\ell} h^{ij} + \left(
\begin{array}[c]{cc}
\dfrac{\gamma(T,R)}{\ell\sinh^4(T/\ell)} & \dfrac{\alpha_0}{\left(R^2 + (\theta_0\ell/2\pi)^2\right) \sinh^4(T/\ell)} \\
\dfrac{\alpha_0}{\left(R^2 + (\theta_0\ell/2\pi)^2\right) \sinh^4(T/\ell)} & \beta(T,R) \coth^2(T/\ell)/\ell^3
\end{array}
\right), \cr
\end{align}
where
\begin{align}
\gamma &= \alpha(T,R) - \sqrt{\alpha(T,R)^2 - \alpha_0^2} \\
\beta &= \frac{ \alpha(T,R) \left(  \sqrt{\alpha(T,R)^2 - \alpha_0^2 } - \alpha(T,R)  \right)-\alpha_0^2}{ \alpha(T,R)^2 \sqrt{\alpha(T,R)^2 - \alpha_0^2} } ,
\end{align}
with
\begin{align}
\alpha(T,R) := \frac{R^2 + (\theta_0 \ell/2\pi)^2 }{\ell^2} \frac{\sinh(2T/\ell)}{2},
\end{align}
where $R$ ranges over $(-\infty, +\infty)$ though we must choose $ \alpha(T,R) > \alpha_0$.  The above canonical data satisfies \eqref{BoundaryConditionsh} with $h_{ij}^{(1)}=0$ so we may readily compute the charges
\begin{eqnarray}
J_0 &=&  \frac{\alpha_0 \ell}{4G} , \cr
K_0 &=& - \frac{1 + (\theta_0/2\pi)^2}{8G} .
\end{eqnarray}

Thus $\alpha_0$, $\theta_0$ are constant on any solution (at least when the lapse and shift have the fall off dictated by~\eqref{BoundaryConditionsh}).  Indeed,  using the Bianchi identities one may show that
$\dot{J}_0 = 0 = \dot{K}_0 $ (as evaluated in one asymptotic region) are precisely the conditions for \eqref{sw} to solve the canonical equations of motion with lapse and shift defined by solving any 3 independent sets of these equations.
The resulting lapse and shift can be chosen to satisfy
\begin{align}
N &= 1+ \frac{2\alpha_0^2 \ell^5}{5 \sinh^2(2T/\ell)R^4} + \order{R}{-5} \\
N^R &= \frac{2\alpha_0^2\ell^3}{5\sinh^3(T/\ell)\cosh(T/\ell)R^3} + \order{R}{-6} \\
N^\theta &= \frac{2 \alpha_0 \ell^2}{3 \sinh^2(T/\ell) R^3} + \order{R}{-2}
\end{align}
in one asymptotic region, though for the foliation defined by \eqref{sw} they then diverge in the second asymptotic region.  It would be interesting to find a more well-behaved foliation of the spinning wormhole spacetime, or perhaps an analytic solution for the full spacetime metric.  Such a solution can presumeably be found by considering the above-mentioned quotients of dS${}_3$ and taking  the size of the commuting rotation to be determined by $\alpha_0$.

In the non-spinning case, it is clear that the above construction may be generalized to quotients by groups with more than one generator.  In analogy with \cite{Aminneborg:1997pz,Aminneborg:1998si}, one may construct quotients for which the $T= (\text{constant})$ surface (and thus $I^+$) is an arbitrary Reimann surface with any number of punctures\footnote{For spheres, the number of punctures must be at least $2$.}, where each puncture describes an asymptotic region.  In particular, one may construct solutions with only a single asymptotic region.  We expect that angular momentum may be added to these solutions as above.  The solution also depends on a choice of internal moduli when the Riemann surface is not a sphere.

\subsection{Comparison with Brown-York methods at future infinity}
\label{comparehyp}

We now compare our charges to those obtained in \cite{Strominger:2001pn,balasubramaniandS} using boundary stress tensors on $I^+$.  We wish to evaluate~\eqref{BYCharge} for $k=-1$.   We begin with the special case $d=3$ for which the counterterm is again simply $(d-2) h^{ij}/\ell \kappa$ which results in a term 
\begin{align}
\Delta^{\prime\prime} \pi^{ij} = \pi^{ij} + \frac{(d-2)}{\ell}  h^{ij},
\end{align}
which matches the $\Delta^\prime \pi^{ij}$ term in \eqref{V2} in the limit $T\rightarrow\infty$.  Thus the $J_n$ agree with~\cite{Strominger:2001pn,balasubramaniandS} on $\CVPSh$.

The situation for $K_n$ is more subtle.  While our $K_n$ are fully gauge invariant, the corresponding Brown-York charges~\eqref{BYCharge} fail to be invariant under all of our gauge transformations.  From the perspective of \cite{Strominger:2001pn,balasubramaniandS}, this is simply because our gauge transformations act as conformal transformations on $I^+$ and thus generate conformal anomaly terms.  These terms are large enough to contribute to the Brown-York versions of the $K_n$  and can even make them diverge.  Under a general transformation that we consider to be gauge, the $K_n$ computed as in \cite{Strominger:2001pn,balasubramaniandS} thus transform by adding a term that depends on the gauge transformation but is otherwise independent of the solution on which it is evaluated; the term is a gauge-dependent $c$-number.  On the other hand, if we impose the gauge \eqref{ggh} we may use \eqref{simple}.  As a result, our $K_n$ charges then match those defined using~\eqref{BYCounterterm} as given by a flat boundary metric.

For $d=4,5$ we will also obtain a $\Delta^{\prime\prime} \pi^{ij} $ term which again reproduces the $\Delta^{\prime} \pi^{ij} $ term in \eqref{Qk=-1} when $T\rightarrow\infty$.  Thus we must again show that the counterterm
\begin{align}
\frac{\ell}{d-3} \mathcal{G}^{ij}
\end{align}
can contribute only a constant (which must then vanish by symmetry for all charges).  Using the same notation and conventions as in section \ref{compare} we first evaluate the contributions from radial momentum constrains. As in the $k=0$ case the contribution is a pure divergence for rotations.  For translations the result is (see Appendix~\ref{details})
\begin{align}
\frac{\hat R_i \ell \mathcal{G}^{ij} \vec{\vec{\zeta}}_j}{d-3} =  \frac{R \vec{\zeta}_\perp}{\ell} \hat{R}^k \sigma^{ij} D_k \Delta \sigma_{ij}. \label{radmomentumhyp}
\end{align}
The contribution from the radial Hamiltonian constraint is given by
\begin{align}
\frac{\hat R_i \ell \mathcal{G}^{ij} \vec{\zeta}_\perp \hat R_j}{d-3} = \frac{R \vec{\zeta}_\perp}{\ell} \hat{R}^k \sigma^{ij}\left( D_i \Delta \sigma_{jk} - D_k \Delta \sigma_{ij} \right). \label{radhamiltonianhyp}
\end{align}
This vanishes explicitly for rotations (for which $\zeta_\perp =0$). For translations, \eqref{radmomentumhyp} nicely cancels the second term in the radial Hamiltonian contribution leaving only
\begin{align}
\frac{\hat R_i \ell \mathcal{G}^{ij} \vec{\zeta}_j}{d-3} = \frac{R \vec{\zeta}_\perp}{\ell} \hat{R}^k \sigma^{ij} D_i \Delta \sigma_{jk}.
\end{align}
The leading order term vanishes because $\zeta_\perp$ is odd.  Power counting now shows that the remaining terms vanish by~\eqref{BoundaryConditionsh}.

\section{Discussion}
\label{disc}

We have constructed phases spaces $\CVPS$ and $\CVPSh$ associated with spacetimes asymptotic to the planar- and hyperbolic-sliced regions $\dS$ and $\dSh$ of de Sitter space for $d \ge 3$.  Our charges agree with those defined in \cite{balasubramaniandS,StromingerASG}  when the latter are computed for our asymptotic Killing fields at the respective $i^0$, and using an appropriate conformal frame for $d=3, k=-1$.  This establishes that (some of) the charges of \cite{Strominger:2001pn,balasubramaniandS,StromingerASG} generate diffeomorphisms in a well-defined phase space.

For $d \ge 4$ our phase spaces are non-trivial in the sense that they contain the de Sitter-Schwarzschild solution as well as spacetimes with generic gravitational radiation through $I^+$, provided only that this radiation falls off sufficiently quickly at $i^0$.  For $d=3$ and $k=0$ the phase spaces become non-trivial when coupled to matter fields or point particles.   Despite the lack of local degrees of freedom, the case $d=3$ with $k=-1$ is non-trivial even without matter due to both boundary gravitons and the family of wormhole spacetimes described in section \ref{wormholes}.  These solutions are $\Lambda > 0$ analogues of BTZ black holes (and their generalizations \cite{Aminneborg:1997pz,Aminneborg:1998si}) and exhibit a corresponding mass gap.  They are labeled by  their angular momentum, an energy-like $K_0$ charge, and the topology of $I^+$, as well as internal moduli if the topology of $I^+$ is sufficiently complicated.  The latter two are analogues of the parameters discussed in \cite{Aminneborg:1997pz,Aminneborg:1998si} for $\Lambda < 0$.

In most cases we found an asymptotic symmetry group (ASG) isomorphic to the isometries of $\dS$ or $\dSh$, though for $k=-1$ and $d=3$ the obvious rotational symmetry was enlarged to a (single) Virasoro algebra in the ASG.  Since we do not include a gravitational Chern-Simons term, the central charge for this case vanishes due to reflection symmetry in the angular direction.  While we expect a similar structure to arise for phase spaces asymptotic to general 2+1 dimensional $k=-1$ Friedmann-Lema\^itre-Robertson-Walker cosmologies, we leave such a general study for future work. We also identified a larger algebra containing two Virasoro sub-algebras with non-trivial imaginary central charges
$\pm i \left(\frac{3\ell}{2G}\right)$ in agreement with those expected from \cite{Maldacena:2002vr,Balasubramanian:2002zh} and computed in \cite{Ouyang:2011fs}.
While the classical reality conditions are those of \cite{Fjelstad:2002wf,Balasubramanian:2002zh,Ouyang:2011fs}, the fact that the additional generators $K_n$ do not preserve our phase space suggests that corresponding ``real'' elements of the algebra (e.g., $K_0$) do not define self-adjoint operators at the quantum level.  Instead, we expect that these operators behave somewhat like $-i \frac{\partial}{\partial x}$ on the half-line and may have complex eigenvalues.  While the $K_n$ charges do not generate symmetries, they are nevertheless gauge invariant and conserved on $\CVPSh$.

A similar extension to the full Euclidean conformal group may also be allowed for $d = 3, k=0$ and perhaps even in higher dimensions.  However, reflection and rotation symmetries imply that the additional `charges' vanish for $d=3$ spinning conical defects, for $d=4$ Kerr-de Sitter, and for rotating de Sitter Myers-Perry black holes in higher dimensions.  We have not investigated whether other solutions in our phase space (perhaps a `moving' black hole?) might lead to non-zero values of such charges.

It remains to compare our phase spaces with those of \cite{Shiromizu:2001bg,KastorCKV,LuoPositiveMass}.  These references studied spacetimes asymptotic to $\dS$ in four spacetime dimensions, so we limit the comparison to our $\CVPS$ with $d=4$.
The focus in \cite{Shiromizu:2001bg,KastorCKV,LuoPositiveMass} was on proving positive energy theorems associated with a charge $Q[\partial_t]$ defined by the time-translation conformal Killing field $\partial_t$ of $\dS$.
Because $\partial_t$ defines only an asymptotic {\it conformal} symmetry, the value of $Q[\partial_t]$ depends on the Cauchy surface on which it is evaluated.  I.e., it is time-dependent, and is thus not conserved in the sense in which we use the term here.  The boundary conditions of~\cite{KastorCKV} for $d=4$ can be stated as follows.  Define
\begin{equation} \label{KTbc}
\Delta^\prime h_{ij} = h_{ij} - e^{2t/\ell} \delta_{ij}= \Delta h_{ij}, \ \ \ \
\Delta^\prime K_{ij} = K_{ij} - \frac{h_{ij}}{\ell} \ne \Delta K_{ij}
\end{equation}
and require that there exist a foliation with vanishing shift on which
$\Delta^\prime h_{ij} = \mathcal{O}(r^{-1})$, $\Delta^\prime K_{ij} = \mathcal{O}(r^{-3})$.
These boundary conditions make $Q[\partial_t]$ finite and allow one to prove that $Q[\partial_t ] \ge0$. However, as the authors note, they do not appear to be sufficient to make finite the charges associated with the asymptotic Killing fields\footnote{One suspects that the boundary conditions of \cite{KastorCKV} can be tightened to make the Killing charges finite while keeping their $Q[\partial_t]$ finite and positive. It would be interesting to show this explicitly.}. In contrast, our boundary conditions were chosen specifically to make such `Killing charges' finite.
Luo et. al.~\cite{LuoPositiveMass} use boundary conditions that are similar to~\cite{KastorCKV}.  They require also that a foliation be constructed with vanishing shift and
$\Delta^\prime h_{ij} = \mathcal{O}(r^{-1})$, $\Delta^\prime K_{ij} = \mathcal{O}(r^{-2})$.

A simple calculation yields
\begin{eqnarray}
\Delta^\prime \tilde{\pi}^{ij} = \Delta \tilde{\pi}^{ij} - \frac{\sqrt{\bar h}}{\ell} \left( 2 \Delta h^{ij} +  \Delta {h^k}_k  \bar h^{ij}\right) + \mathcal{O}(\Delta h^2).
\end{eqnarray}
Now in order for Eqs.~\eqref{BoundaryConditions} and the boundary conditions of either \cite{KastorCKV} or \cite{LuoPositiveMass} to be simultaneously satisfied, we must have at least $\Delta h_{ij} = \Delta^\prime h_{ij} = \mathcal{O}(r^{-2})$.  Using the results of~\cite{LuoPositiveMass} (particularly Theorem 4.2) it can be shown that the only globally hyperbolic spacetime which satisfies both sets of boundary conditions on the same foliation is exact de Sitter space in the form \eqref{dsHypVacuum} (up to gauge transformations).\footnote{However it is possible for the same spacetime to admit two different foliations with one satisfying \eqref{BoundaryConditionsh} and the other satisfying the boundary conditions of \cite{KastorCKV,LuoPositiveMass}.  This is in particular the case for the de Sitter--Schwarzschild solution; see \cite{KastorCKV,LuoPositiveMass}.}  Our phase space thus has precisely one point in common with that of either \cite{KastorCKV} or \cite{LuoPositiveMass}.  It is nevertheless interesting to ask whether the methods of \cite{KastorCKV,LuoPositiveMass} might be used to derive a bound on the charge $K_0$ for $d=3$.

We end with a brief comment on the related approach of Shiromizu et. al.~\cite{Shiromizu:2001bg}, which also imposes $\Delta^\prime h_{ij} = \mathcal{O}(r^{-1})$, $\Delta^\prime K_{ij} = \mathcal{O}(r^{-3})$ and in addition requires the sapcetime to admit a foliation by constant mean curvature slices with $h^{ij} \Delta^\prime K_{ij} = 0$.  This last requirement is quite non-trivial, though it has been shown numerically that it continues to allow Schwarzschild-de Sitter black holes~\cite{Nakao:1990gw}.  Results from the asymptotically flat case~\cite{Isenberg:1997re,FischerMoncrief1997} suggest that constructing a phase space of such solutions is non-trivial, but possible with the right understanding.  Due to the similarities with \cite{KastorCKV,LuoPositiveMass}, we expect that our charges would again fail to be finite on such a phase space, but this remains to be shown in detail.

\section*{Acknowledgements}
We thank Tomas Andrad\'e, Curtis Asplund, Andy Strominger, Jennie Traschen, and David Kastor for interesting discussions concerning de Sitter charges.  We also thank Alejandra Castro, Matthias Gaberdiel, and Alex Maloney for discussions related to the algebra \eqref{ccharge}.  Finally we thank an anonymous referee for identifying an error in a previous draft.  This work was supported in part by the National Science Foundation under Grant No PHY08-55415, and by funds from the University of California.

\appendix

\section{Finiteness of the charges for $\CVPS$.}
\label{finiteQ}

As noted in section \ref{CC}, despite naive power-counting divergences in \eqref{Hz},
finiteness of the symplectic structure \eqref{SymplecticStructure} under the boundary conditions \eqref{BoundaryConditions} guarantees that charges defined by asymptotic symmetries are in fact finite on solutions.
We now show this explicitly by solving the constraints at the leading orders in $1/r$.  Below, we use the notation defined in \eqref{2ndparts}.

First we show that the charges are given by~\eqref{Hk=0}.  The second surface integral in~\eqref{Hz} vanishes by power counting.  As for the first surface integral, note that the following two terms combine nicely
\begin{align}
\int_{\partial\Sigma} (dr)_i \vec{\xi}^j \left( \Delta \tilde\pi^{ik} h_{kj} - \frac{\tilde\pi^{kl} \Delta h_{kl} }{2} {\delta^i}_j \right) &= \int_{\partial\Sigma} \sqrt{\bar h}  (dr)_i \vec{\xi}^j \left( \Delta \pi^{ik} h_{kj} + \frac{  \bar h^{lm} \Delta h_{lm} }{2} {\bar \pi^i}_j - \frac{ \bar \pi^{kl} \Delta h_{kl} }{2} {\delta^i}_j \right) \nonumber \\
&= \int_{\partial\Sigma} \sqrt{\sigma}  \vec{\xi}^j  \Delta \pi^{rk} \bar h_{kj}, \label{traceterm}
\end{align}
where we have used~\eqref{BoundaryConditions} and simple power counting to drop subleading terms.  With this result,~\eqref{Hz} becomes
\begin{align} \label{Hfinite}
H(\xi) &= \frac{1}{\kappa} \int_{\partial\Sigma} \sqrt{\sigma} \vec{\xi}^j  \left(   \Delta \pi^{rk} \bar h_{kj} + \bar \pi^{rk} \Delta h_{kj} \right) \cr
&= \frac{1}{\kappa} \int_{\partial\Sigma} \sqrt{\sigma} \vec{\xi}^j  \left[   \pi^{rk}  + \frac{d-2}{\ell} \left( \bar h^{ik} - \Delta h^{ik}  \right) \right] \bar h_{kj} \cr
&=  \frac{1}{\kappa} \int_{\partial\Sigma} \sqrt{\sigma} \vec{\xi}^j  \left(\pi^{rk}  + \frac{(d-2) h^{ik} }{\ell}   \right) \bar h_{kj},
\end{align}
from which we obtain~\eqref{Hk=0}.

Now, we show explicitly that these charges are finite.  We define
\begin{align}
\label{chi0}
\chi_{0}^{ij} &:= \pi_{(d-2)}^{ij}  \\
\chi_{1}^{ij} &:= \pi_{(d-1)}^{ij} + \bar{\pi}^{ik} { {h^{(d-1)}}_k}^{j},
\end{align}
and evaluate the constraints to order $r^{-(d-1)}$,
\begin{align}
D_j \chi_{0}^{ij} &= 0 = \chi_{0} \label{constraint0} \\
D_j \chi_{1}^{ij} &= 0 = \chi_{1}.\label{constraint1}
\end{align}
Using these constraints and the known $r$ dependence of $\chi_0^{ij}$ we find
\begin{align}
\label{chiid}
\chi_0^{ir} &= - \mathcal{D}_{\underline k} \left( r \chi_{0}^{i \underline k}\right).
\end{align}
Now again using~\eqref{traceterm} we can write~\eqref{Hz} as
\begin{align} \label{Hchi}
H(\xi) &=  \frac{1}{\kappa} \int_{\partial\Sigma}\sqrt{\sigma} \left( \chi_{0}^{r k}+ \chi_{1}^{r k}  \right)\bar h_{kj} \vec\xi^j  .
\end{align}
By power counting this expression is finite for translations.  For dilations and rotations the second term is finite by power counting and the first term vanishes.  This can be seen by using \eqref{ckvf}, \eqref{chi0}, \eqref{chiid} and integrations by parts, which give
\begin{align} \label{Hchifirstterm}
\int_{\partial\Sigma} \sqrt{\bar \sigma}  \chi_{0}^{r k} \bar{h}_{kj}  \vec{\xi}^j &= \int_{\partial\Sigma} \sqrt{\bar \sigma} \left( \chi_{0}^{rr} \vec{\xi}_\perp + \chi_{0}^{r \underline i } \vec{\vec{\xi}}_{\underline{i}} \right) \\
&= \int_{\partial\Sigma} \sqrt{\bar \sigma}  \left(2 \chi_{0}^{rr} \vec{\xi}_\perp   \right) \\
&= \int_{\partial\Sigma} \sqrt{\bar \sigma}  \left( 2 r\chi_{0}^{r\underline j} \mathcal{D}_{\underline j}\vec{\xi}_\perp   \right) = 0.
\end{align}

From~\eqref{Hchi} and~\eqref{Hchifirstterm} we can see that solutions to~\eqref{ckvf} which vanish on $\partial\Sigma$ are gauge transformations as follows:  Any such solution $\xi$ has a Hamiltonian $H(\xi)$ which is identically zero.  Using the identity $\omega(\delta g, \pounds_\xi g)=\delta H(\xi)$ where $\delta g$ is an arbitrary tangent vector, we see that $\pounds_\xi g$ is a degenerate direction of the symplectic structure for such $\xi$.  Thus any vector $\xi$ which preserves our boundary conditions and vanishes at infinity generates a gauge transformations.

\section{The radial Hamiltonian constraint}

\label{details}

We now fill in some details of the analysis of the radial Hamiltonian constraint \eqref{rH} in sections~\ref{compare} and~\ref{comparehyp}.

\subsection{$k=0$}

Let us write $\sigma_{ij} = \bar \sigma_{ij} + \Delta \sigma_{ij}, \theta_{ij} = \bar \theta_{ij}+\Delta\theta_{ij}$, where $\bar \sigma_{ij}$, $\bar \theta_{ij}$ are the induced metric and extrinsic curvature of $\partial \Sigma_\infty$ in  exact $\dS$.  In particular,
\begin{subequations}\label{bg}
\begin{align}
\bar \theta_{ij} &= \frac{\bar \sigma_{ij}}{r} \\
\bar \sigma^{ij}\Delta\theta_{ij} &= - \hat{r}^m \bar \sigma^{jk} \left(D_j \Delta h_{km} - \frac{1}{2} D_m \Delta h_{jk}\right) + \left(\textrm{Pure Divergence on $\partial \Sigma_\infty$}\right) \\
\bar{\mathfrak{R}} &= (d-3) \frac{\bar{\sigma}_{ij}}{r^2}.
\end{align}
\end{subequations}
Thus
\begin{eqnarray}
\Delta \left(-\theta^2 + \theta_{ij}\theta^{ij}\right) &=&   \frac{d-3}{r} \left[-\frac{2  \bar \sigma_{ij}\Delta\sigma^{ij}}{r} + \hat{r}^m \bar \sigma^{jk} \left(2D_j \Delta h_{km} -  D_m \Delta h_{jk}\right)\right] + \dots, \cr
\Delta \mathfrak{R}_{ij} &=&  \frac{d-3}{r}\frac{\Delta \sigma^{ij} \bar \sigma_{ij}}{r} + \dots.
\end{eqnarray}
where $\dots$ represents a linear combination of higher order terms in $\Delta h_{ij}$ and  total divergences on $\partial \Sigma_\infty$.

Next we use the definition of $\theta_{ij}$ to note that
\begin{equation}
\hat{r}^m \sigma^{jk}  D_j \Delta h_{km} =  -    \Delta h_{km} \theta^{km} + \left(\textrm{Pure Divergence on $\partial \Sigma_\infty$}\right).
\end{equation}
Using \eqref{bg} then yields
\begin{equation} \label{ApIdentity}
\hat{r}^m \sigma^{jk}  D_j \Delta h_{km} = -   \frac{\Delta \sigma_{km}\sigma^{km}}{r} + \dots =  \frac{\Delta \sigma^{km}\sigma_{km}}{r} + \dots.
\end{equation}
Combining the results above gives (\ref{pp}).

\subsection{$k=-1$}

With the same notation as above, for $k=-1$ we have
\begin{subequations}\label{bghyp}
\begin{align}
\bar \theta_{ij} &= \frac{R \bar \sigma_{ij}}{\ell^2} \\
\bar \sigma^{ij}\Delta\theta_{ij} &= - \hat{R}^m \bar \sigma^{jk} \left(D_j \Delta h_{km} - \frac{1}{2} D_m \Delta h_{jk}\right) + \left(\textrm{Pure Divergence on $\partial \Sigma_\infty$}\right) \\
\bar{\mathfrak{R}} &= (d-3) \frac{\bar{\sigma}_{ij}}{r^2},
\end{align}
\end{subequations}

To derive~\eqref{radmomentumhyp} we use the fact that the translation symmetries are conformal Killing vectors on $\partial\Sigma$ which satisfy
\begin{align}
\mathcal{D}_{(i}\vec{\vec{\zeta}}_{j)} = -\frac{R\vec{\zeta}_\perp}{\ell^2}\sigma_{ij} + \dots,
\end{align}
so
\begin{align}
\Delta\left(\mathcal{D}_i(\theta^{ij}-\theta\sigma^{ij})\vec{\vec{\zeta}}_j\right) &= \Delta\left((\theta^{ij}-\theta\sigma^{ij})\mathcal{D}_{(i}\vec{\vec{\zeta}}_{j)} + \dots \right) \\
&=  \frac{(d-3)R \vec{\zeta}_\perp}{2 \ell^2} \hat{R}^m \bar \sigma^{jk} D_m \Delta \sigma_{jk} + \dots.
\end{align}

As for the radial Hamiltonian constraint, we have
\begin{eqnarray}
\Delta \left(-\theta^2 + \theta_{ij}\theta^{ij}\right)  &=&   \frac{(d-3)R}{\ell^2} \left[-\frac{2  R \bar \sigma_{ij}\Delta\sigma^{ij}}{\ell^2} + \hat{R}^m \bar \sigma^{jk} \left(2D_j \Delta h_{km} -  D_m \Delta h_{jk}\right)\right] + \dots, \cr
\Delta \mathfrak{R} &=& \frac{(d-3)R}{\ell^2} \frac{R \Delta \sigma^{ij} \bar \sigma_{ij}}{\ell^2} + \dots,
\end{eqnarray}
which, after using~\eqref{ApIdentity} to combine terms, gives~\eqref{radhamiltonianhyp}.

\section{ASG of $\CVPSh$ for $d=3$}
\label{d=3ASG}

For $d=3$, the solutions to \eqref{ckvfHyp} are given by
\begin{align}
\xi_\theta &= \sum_n e^{in\theta} R f_n(T,R) \\
\xi_R &= \sum_n \frac{e^{in\theta}}{ i n } \left(2 f_n(T,R) -\frac{f_n^\prime(T,R)}{R}\right) \\
\xi_T &= \frac{1+R^2/\ell^2}{\sinh(T/\ell)\cosh(T/\ell)} \left( \ell  \xi^\prime_R + \frac{R}{\ell} \xi_R \right),
\end{align}
where primes signify $R$ derivatives and $f_n$ is the solution to
\begin{align}
R^2 (1+R^2/\ell^2) f^{\prime\prime}_n - R f^\prime_n + (1-n^2) f_n = 0.
\end{align}

Using the ansatz
\begin{align}
f_n(R,T) = \sum_{k=-\infty}^\infty \epsilon_n^{(k)}(T) R^k,
\end{align}
we find that $\epsilon_n^{(k)}$ must satisfy the recursion relation
\begin{align} \label{recursionrelation}
\left[ (k-1)^2 - n^2 \right] \epsilon_n^{(k)} = - \frac{(k-2)(k-3)}{\ell^2} \epsilon_n^{(k-2)}.
\end{align}
From this relation we see that $\epsilon_n^{(k\ge2)} = 0$ and the solution has two independent integration constants $\epsilon_n^{(1)}$ and $\epsilon_n^{(0)}$, from which the rest of the series is determined (though as shown in appendix \ref{finiteQ} the terms involving $\epsilon_n^{(k < 0)}$ are pure gauge).

Finally, we must specify the time dependence of $\epsilon_n^{(1)}$ and $\epsilon_n^{(0)}$.  We will use this freedom to enforce $\bar \nabla_{(T} \xi_{i)}\sim \order{R}{-2}$.  This condition is met by
\begin{align}
\epsilon_n^{(1)} &= A_n \sinh^2(T/\ell) \\
\epsilon_n^{(0)} &= B_n \sinh(T/\ell)\cosh(T/\ell),
\end{align}
which gives
\begin{align}
\zeta_n^a = e^{in\theta}&\left(A_n + B_n \frac{\coth(T/\ell)}{R}\right)(\partial_\theta)^a + \frac{e^{in\theta}}{in}\left(B_n \frac{R^2 \coth(T/\ell)}{\ell^2} + A_n n^2 R\right)(\partial_R)^a \cr
&+  \frac{e^{in\theta}}{in}\left(-B_n \left(\frac{R}{\ell}-\frac{\ell(n^2-1)}{2R}\right) + A_n \frac{n^2(n^2-1)\ell^3}{3R^2\coth(T/\ell)}\right)(\partial_T)^a+\dots ,
\end{align}
where here and below $\dots$ denote pure gauge terms.

Now we define $\xi_{K_n}$ by $A_n = 0 $ and $B_n = -in $ and $\xi_{J_n}$ by $A_n = 1$ and $B_n = 0$
\begin{align}
\xi_{K_n} &=  e^{in\theta} \left( \frac{R}{\ell} \partial_T - \frac{R^2 \coth(T/\ell)}{\ell^2} \partial_R - \frac{in\ell \coth(T/\ell)}{R} \partial_\theta + \dots \right), \\
\xi_{J_n} &= e^{in\theta} \left( - \frac{in(n^2-1) \ell^3}{3 R^2 \coth(T/\ell)} \partial_T -i n R \partial_R +  \partial_\theta + \dots \right).
\end{align}

The charges $L_n$ ($\bar L_n$) are now given by
\begin{align}
L_n &= \frac{J_n + i \ell K_n}{2} \cr
\bar L_n &= \frac{J_n - i \ell K_n}{2},
\end{align}
which lead to the algebra \eqref{ccharge}.

\section{Gauge Invariance, Finiteness, and Conservation of $K_n$ in $\CVPSh$}
\label{KnConservation}

\subsection{Gauge Invariance}

Recall that expression \eqref{V2} for $K_n$ is valid only when $h^{(1)}_{ij} =0$.  Let us first show that \eqref{V2} is invariant under the remaining gauge transformations.  Invariance under spatial diffeomorphisms is manifest, so it remains only to consider diffeomorphisms generated by $\zeta = \zeta_\perp \partial_T$. The boundary conditions of $\CVPSh$ and the condition $h^{(1)}_{ij} =0$ require that $\zeta_\perp = \zeta_\perp^{(2)}(T,\theta) R^{-2} + \order{R}{-3}$.  Thus we have
\begin{align}
\delta_\zeta \Delta h_{ij} &=   \zeta_\perp \partial_T \bar h_{ij} + \dots \cr
&= 2 \frac{ \zeta_\perp^{(2)}}{R^2} \frac{\coth(T/\ell)}{\ell}  \bar h_{ij} + \dots,
\end{align}
where $\dots$ indicate terms which fall off too fast too contribute to~\eqref{V2}.  The behavior of $\pi^{ij}$ under such a transformation is more complicated, however a straightforward computation gives
\begin{align}
\delta_\zeta \pi^{RR} =  \zeta_\perp^{(2)} \frac{ [4+\cosh(2T/\ell)] \csch^4(T/\ell) }{\ell^4} + \dots .
\end{align}
Inserting these result into~\eqref{V2} ultimately gives $\delta_\zeta K_n=0$.
Having shown \eqref{V2} to be invariant under gauge transformations preserving the condition $h^{(1)}_{ij} =0$, we may extend the definition of $K_n$ to the full phase space by taking it to be fully gauge invariant.

\subsection{Finiteness}

Since $K_n$ is gauge invariant we now restrict to solutions satisfying the condition \eqref{ggh} so that we may use the simple expression \eqref{simple}. 
First we must show that this is finite.  We have 
\begin{align}
K_n &= \frac{1}{\kappa} \int_{\partial\Sigma} \sqrt{\sigma} \hat{R}_i \Delta \pi^{ik} h_{kj}  \frac{R^2\coth(T/\ell)}{\ell^2} (\partial_R)^j e^{in\theta} \cr
&= \frac{1}{\kappa} \int_{\partial\Sigma} (\pi_{(1)}^{RR} + \pi_{(2)}^{RR}) \sinh^3(T/\ell)\cosh(T/\ell) (\partial_R)^j e^{in\theta}.
\end{align}
Imposing the constraints on the boundary conditions of $\CVPSh_{gf}$ we find that
\begin{align}
D_i \pi^{ij}_{(1)} &= 0 = \pi_{(1)} \cr
D_i \pi^{ij}_{(2)} &= 0 = \pi_{(2)}.
\end{align}
Using these expressions we can show that
\begin{align}
{\cal D}_{\underline i} \pi^{\underline{i}R}_{(1)} &=  -\frac{\pi^{RR}_{(1)}}{R} \cr
{\cal D}_{\underline i} \pi^{\underline{i}R}_{(2)} &= 0.
\end{align}
Thus we see that the $\pi_{(1)}^{RR}$ term is a pure divergence which vanishes upon integration over the sphere.  The remaining term is finite by power counting.

For future reference we note that the constraints also require
\begin{align}
\pi_{(2)}^{\theta\theta} &= - \frac{\ell^2\pi_{(2)}^{RR}}{R^4},
\end{align}
so $\pi_{(2)}^{ij}$ only has two independent components.

\subsection{Conservation}

In the gauge $h_{ij}^{(1)} = h_{ij}^{(2)} = 0$ the equation of motion for the induced metric is
\begin{align}
\dot{\bar{h}}_{ij} = 2 N (\pi_{ij} - \pi \bar{h}_{ij}) + 2 \bar{D}_{(i} N_{j)} + \dots,
\end{align}
where $\dots$ represents terms that fall off like $h^{(3)}_{ij}$ or faster.  Solving this equations to lowest order and applying the constraints gives
\begin{align}
N &= 1 + \frac{\ell^3 \sinh^2(T/\ell) \tanh(T/\ell) \pi_{(2)}^{RR}}{3 R^2} + \dots \cr
N^R &= \frac{2 \ell^2 \sinh^2(T/\ell) \pi_{(2)}^{RR} }{3 R} + \dots \cr
N^\theta &= \frac{2 \ell^3 \sinh^2(T/\ell) \pi_{(2)}^{R\theta} }{2 R^3} + \dots,
\end{align}
which satisfy~\eqref{BoundaryConditionsh}.  Inserting these expressions into the equation of motion for the momentum and again applying the constraints gives two first order differential equations for the two independent components of $\pi_{(2)}^{ij}$
\begin{align}
0 &= \ell \dot\pi_{(2)}^{RR} + 2\left( 2\coth(2T/\ell)+\csch(2T/\ell) \right) \pi_{(2)}^{RR} \cr
0 &= \ell \dot\pi_{(2)}^{R\theta} + 4 \coth(T/\ell) \pi_{(2)}^{R\theta}.
\end{align}
The solutions  are
\begin{align}
\pi_{(2)}^{RR} &= \frac{\kappa}{2\pi \sinh^3(T/\ell) \cosh(T/\ell)} K(\theta)\cr
\pi_{(2)}^{R\theta} &= \frac{\kappa}{2\pi \sinh^4(T/\ell) R^2} J(\theta) ,
\end{align}
where $K(\theta)$ and $J(\theta)$ are free functions that depend only on $\theta$.  Comparison with \eqref{simple} then shows that $K_n,J_n$ are the Fourier components of $K(\theta), J(\theta)$ and thus are conserved on $\CVPSh$.  This is the desired result for $K_n$.  For $J_n$ (which form the ASG), it is a simple check that our phase space is well-defined.

\bibliography{deSitterChargesArxivV2.bbl}

\end{document}